\documentclass[twocolumn,superscriptaddress,preprintnumbers,amsmath,amssymb]{revtex4}
\usepackage{amsmath}
\usepackage{amssymb}
\usepackage{dcolumn}% Align table columns on decimal point
\usepackage{graphicx}
\usepackage{bm}% bold math
\usepackage{epstopdf}
\usepackage{threeparttable}
\usepackage{float}
\usepackage{color}
\usepackage{cleveref}

\usepackage{appendix}

\usepackage [english]{babel}
\usepackage [autostyle, english = american]{csquotes}
\MakeOuterQuote{"}
\usepackage[normalem]{ulem}%cross-out

\makeatletter
\newcommand*{\rom}[1]{\expandafter\@slowromancap\romannumeral #1@}
\makeatother
\makeatletter
\def\p@subsection{}
\makeatother

\begin{document}

%\preprint{}

%\title{Atomic rearrangement: onset of energy dissipation in vibrating metallic glass nano-resonators / Intrinsic dissipation mechanism and quality factor quantification in model glasses.} %Title of paper
\title{Intrinsic dissipation mechanisms in metallic glass resonators}



\author{Meng Fan}
\affiliation{Department of Mechanical Engineering and Materials Science, Yale University, New Haven, Connecticut, 06520, USA}
\affiliation{Center for Research on Interface Structures and Phenomena, Yale University, New Haven, Connecticut, 06520, USA}
\author{Aya Nawano} 
\affiliation{Department of Mechanical Engineering and Materials Science, Yale University, New Haven, Connecticut, 06520, USA}
\affiliation{Center for Research on Interface Structures and Phenomena, Yale University, New Haven, Connecticut, 06520, USA}
\author{Jan Schroers}
\affiliation{Department of Mechanical Engineering and Materials Science, Yale University, New Haven, Connecticut, 06520, USA}
\affiliation{Center for Research on Interface Structures and Phenomena, Yale University, New Haven, Connecticut, 06520, USA}
\author{Mark D. Shattuck}
\affiliation{Department of Physics and Benjamin Levich Institute, The City College of the City University of New York, New York, New York, 10031, USA}
\affiliation{Department of Mechanical Engineering and Materials Science, Yale University, New Haven, Connecticut, 06520, USA}
\author{Corey S. O'Hern}
\affiliation{Department of Mechanical Engineering and Materials Science, Yale University, New Haven, Connecticut, 06520, USA}
\affiliation{Center for Research on Interface Structures and Phenomena, Yale University, New Haven, Connecticut, 06520, USA}
\affiliation{Department of Physics, Yale University, New Haven, Connecticut, 06520, USA}
\affiliation{Department of Applied Physics, Yale University, New Haven, Connecticut, 06520, USA}


%\date{\today}

\begin{abstract}
Micro- and nano-resonators have important applications including
sensing, navigation, and biochemical detection. Their performance is
quantified using the quality factor $Q$, which gives the
ratio of the energy stored to the energy dissipated per cycle.
Metallic glasses are a promising materials class for micro- and
nano-scale resonators since they are amorphous and can be fabricated
precisely into complex shapes on these lengthscales. To understand the
intrinsic dissipation mechanisms that ultimately limit large
$Q$-values in metallic glasses, we perform molecular dynamics
simulations to model metallic glass resonators subjected to bending
vibrations.  We calculate the vibrational density of states,
redistribution of energy from the fundamental mode of vibration, and
$Q$ versus the kinetic energy per atom $K$ of the excitation.  In the
linear and nonlinear response regimes where there are no atomic
rearrangements, we find that $Q \rightarrow \infty$ (since we do not
consider coupling to the environment).  We identify a characteristic
$K_r$ above which atomic rearrangements occur,
and there is significant energy leakage from the fundamental mode to
higher frequencies, causing finite $Q$.  Thus, $K_r$ is a critical
parameter determining resonator performance. We show that $K_r$
decreases as a power-law, $K_r\sim N^{-k},$ with increasing system
size $N$, where $k \approx 1.3$.  We estimate the critical strain
$\langle \gamma_r \rangle \sim 10^{-8}$ for micron-sized resonators
below which atomic rearrangements do not occur, and thus large
$Q$-values can be obtained when they are operated below $\gamma_r$.
We find that $K_r$ for amorphous resonators is comparable to that for
resonators with crystalline order.
\end{abstract}

%\pacs{}% insert suggested PACS numbers in braces on next line

\maketitle %\maketitle must follow title, authors, abstract and \pacs

\section{Introduction} 
\label{sec:intro}

Micro- and nano-resonators have numerous important applications
including navigation, sensing, chemical detection, molecular
separation, and biological imaging~\cite{arash2015review}. The
performance of resonators is typically measured by the quality factor,
$Q$, which gives the ratio of the energy stored to the
energy dissipated per cycle in the resonator during
operation~\cite{green1955story}. Micro- and nano-resonators made from
non-metallic crystalline materials, such as
sapphire~\cite{tobar1998high}, carbon
nanotubes~\cite{huttel2009carbon,jiang2004intrinsic}, and single-crystal
diamond~\cite{ovartchaiyapong2012high} can possess quality factors
$Q > 10^6$ at low temperatures. However, it is difficult
to fabricate these materials into complex shapes, and many applications
require electrical conduction.  As a result, crystalline metals are 
used in many resonator applications, yet these suffer from energy
losses that arise from topological defects and grain
boundaries~\cite{blanter2007internal}.

In an effort to obviate energy losses from topological defects and
grain boundaries that occur in crystalline metals, as well as take 
advantage of their plastic-forming ability to be fabricated into 
complex shapes, several groups have considered
resonators made from metallic glasses (MGs)~\cite{khonik1996nature,kanik2014high,kanik2015metallic,hiki2009internal,bardt2007micromolding}. MGs are cooled rapidly to avoid crystallization, and thus they
possess uniformly disordered structure.  Recent experiments have shown
that metallic-glass-based resonators can achieve quality factors that are
comparable and even larger than those for resonators made from
crystalline metals~\cite{kanik2014high}. Metallic glasses offer the
additional benefit for resonator applications in that they can be
thermoplastically formed into complex shapes with spatial features
that span many orders of
magnitude~\cite{schroers2010processing,kumar2009nanomoulding,li2018atomic}.  In
this work, we seek to characterize the dissipation mechanisms that
determine the quality factor for metallic glass resonators.

The mechanisms that give rise to energy losses during vibration can be
classified as intrinsic or extrinsic~\cite{arash2015review}. Extrinsic
losses, such as anchoring and frictional losses, come from
interactions between the resonator and its surrounding
environment~\cite{perez2007design,shkel2006type}.  In contrast,
intrinsic losses originate from flaws or defects within the resonator,
such as dislocations, grain boundaries, vacancies, and interstitials
in crystalline materials. In metallic glasses, which lack crystalline
order, intrinsic losses are envisioned to stem from irreversible,
collective atomic rearrangements, or shear-transformation zones
(STZs)~\cite{falk1998dynamics}.  A number of studies have
characterized the role of collective atomic rearrangements in
determining the mechanical properties of metallic glasses, including
ductility, yielding, and shear-band
formation~\cite{fan2017effects,yu2012tensile,fan2017particle,zemp2015crystal,ketkaew2018mechanical}.

Internal friction measurements have been performed to gain insight
into intrinsic dissipation mechanisms and the quality factor of
metallic
glasses~\cite{zener1937internal,nowick2012anelastic,blanter2007internal}. However,
a key focus in this work has been on revealing structural relaxation
processes at elevated temperatures, rather than their room temperature
behavior (which is typically significantly below the glass transition
temperature). In these studies, metallic glass samples are typically
perturbed by mechanical or electrostatic excitation using a torsion
pendulum or dynamical mechanical analyzer, and the internal friction
is measured as a function of temperature, frequency, and strain
amplitude~\cite{barmatz1974young,hiki2008temperature,blanter2007internal}.
The internal friction, which is proportional to $Q^{-1}$, is generally
small for temperatures below room temperature, and then increases
dramatically, forming a strong peak at temperatures typically above
$400$-$500$~K due to collective $\alpha$ structural
relaxations~\cite{sinning1988influence,samwer1995dynamic,khonik1996nature}.
Studies~\cite{blanter2007internal} have also reported a much smaller
peak (typically four orders of magnitude smaller than that
corresponding to $\alpha$ relaxations) in the internal friction of
metallic glasses between $50$~K and room temperature. Researchers have
suggested that this peak corresponds to localized, anelastic so-called
$\beta$ relaxations. Explanations of the peaks in the internal
friction include the creation and destruction of free
volume~\cite{spaepen1977microscopic}, dislocation
motion~\cite{gilman1973flow}, shear transformation
zones~\cite{falk1998dynamics}, shear bands, and other mechanisms that
involve structural rearrangements.  At even lower temperatures
($<50$~K), the internal friction has been described using the quantum
mechanical tunneling model for two-level
systems~\cite{anderson1972anomalous,phillips1972tunneling}.

Most of these prior studies of the vibrational properties of metallic
glasses either use a quantum mechanical approach for the low-temperature
behavior or consider temperatures near room temperature and above,
where thermal fluctuations are significant and microscopic
rearrangements of atoms are frequent.  In this work, we will take a
different, but still classical approach, and focus on the nearly
zero-temperature regime, where even microscopic rearrangements of
atoms are rare, to better understand the transition from the linear
response regime where $Q$ is infinite (since we do not consider
coupling of the system to the environment) to the highly nonlinear regime
where $Q$ becomes finite.

We carry out molecular dynamics (MD) simulations to
quantify the intrinsic dissipation caused by atomic rearrangements and
measure the quality factor in model metallic glass resonators.  We
induce vibrations in a thin bar-shaped resonator by exciting the mode
corresponding to the resonator's fundamental frequency with a given
kinetic energy per atom $K$ and then running MD at constant total
energy. When $K$ is small, i.e.\ $K<K_{nl}$, the resonator displays
linear response, the spectrum of the vibrational modes only includes the
fundamental mode, and $Q\rightarrow \infty$.  For intermediate $K$,
i.e. $K_{nl}<K<K_r$, energy leaks to modes other than the fundamental
mode, but at sufficiently long times the leakage stops. Thus, in this
regime, $Q \rightarrow \infty$ at long times. For $K > K_r$,
the system undergoes one or more atomic rearrangements, which induce
strong dissipation and finite $Q$. Thus, the magnitude of $K_r$
controls the performance of metallic glass resonators. We further show that
$K_r$ can be increased by decreasing the system size or by decreasing
the cooling rate used to prepare the resonator. We also show that resonators
with amorphous structure can achieve comparable performance (e.g. 
same $Q$) to those with partial crystalline order. 

% structure of paper
The remainder of the article is organized as follows.  In
Sec.~\ref{sec:methods}, we describe the simulation methods we use to
prepare and excite the metallic glass resonators, and to quantify the
energy loss and quality factor of the vibrations.  In
Sec.~\ref{sec:results}, we present the results, including measurements
of the intrinsic loss and dissipation arising from nonlinearity and
atomic rearrangements, techniques to increase $Q$ by decreasing the
system size and cooling rate, and comparisons of resonator performance in
amorphous and crystalline samples. In Sec.~\ref{sec:conclusion}, we
summarize our findings and present promising directions for future
research.  We also include two Appendices. In Appendix A, we show that
our results for the vibrational response do not depend strongly on the
length of the time series of the vibrations that we collect.  In
Appendix B, we show the time dependence of the vibrational density of
states, which supports the findings presented in the main text.

\section{models and methods} 
\label{sec:methods}

\begin{figure}
\begin{center}
\includegraphics[width=0.99\columnwidth]{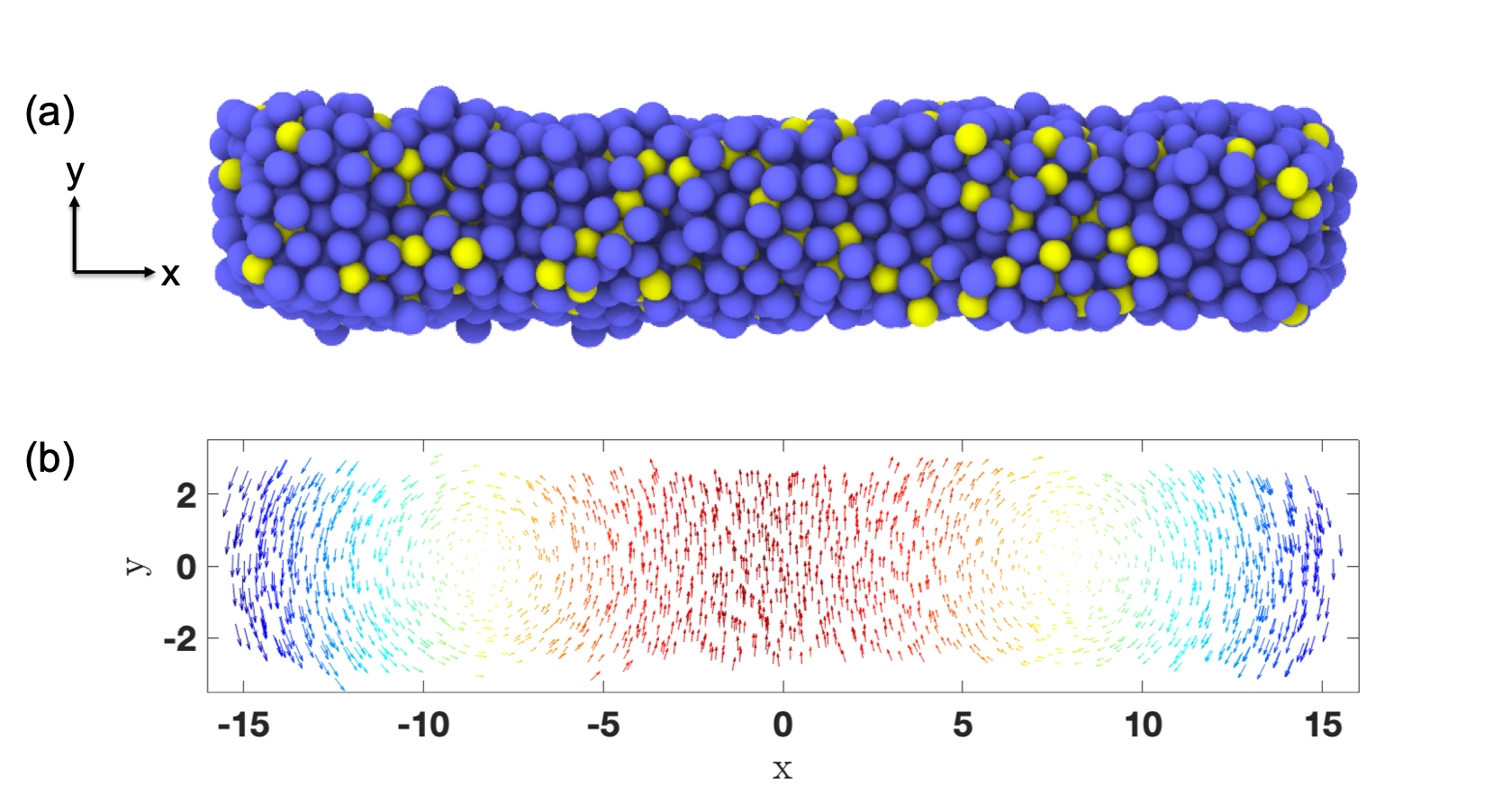}
\caption{(a) View of the model metallic glass resonator along the $z$ axis. 
The bar contains $N=2000$ atoms with aspect ratio $L_x:L_y:L_z = 6:1:2$.
Blue and yellow atoms indicate $A$ and $B$ atom types,
respectively. (b) Vector field representing the fundamental mode of the 
dynamical 
matrix of the metallic glass resonator in (a). The color scale highlights 
the y-component of the fundamental mode contribution for each atom with 
red corresponding to positive and blue corresponding to negative $y$-values.}
\label{fig:method}
\end{center}
\end{figure}

% Method general
We perform molecular dynamics (MD) simulations of binary Lennard Jones
mixtures using the Kob-Andersen model~\cite{kob1995testing}, which has 
been employed to describe NiP alloys. Spherical
atoms interact pairwise via the shifted-force version of the
Lennard-Jones potential, $u(r_{ij}) = 4
\epsilon_{ij}[(\sigma_{ij}/r_{ij})^{12}-(\sigma_{ij}/r_{ij})^6]$ with
a cutoff distance $r_c =2.5 \sigma_{ij}$, where $r_{ij}$ is the
separation between atoms $i$ and $j$. The total potential energy per 
atom is $U = N^{-1} 
\sum_{i>j} u(r_{ij})$. $80\%$ of the atoms are type A
($N_A/N=0.8$) and $20\%$ are type B ($N_B/N=0.2$), where $N=N_A + N_B$ 
is the total number of atoms, and the energy and
length parameters are given by $\epsilon_{AA}=1.0$,
$\epsilon_{BB}=0.5$, $\epsilon_{AB}=1.5$, $\sigma_{AA}=1.0$,
$\sigma_{BB}=0.88$, and $\sigma_{AB}=0.8$.  All atoms have the same
mass $m$. The energy, length, and pressure scales are given in terms of
$\epsilon_{AA}$, $\sigma_{AA}$, and $\epsilon_{AA}/\sigma_{AA}^3$, 
respectively.

% Initialization and cooling.
We initially placed the $N$ atoms on an FCC lattice in a long, thin
box with aspect ratio $L_x:L_y:L_z = 6:1:2$ and periodic boundaries in
the $x$-, $y$-, and $z$-directions at reduced number density $\rho =
1.0$.  We then equilibrated the system at high temperature $T_0 > T_g
\sim 0.4$~\cite{wittmer2013shear} (which melts the crystal) by running
molecular dynamics simulations at fixed number of atoms, pressure, and
temperature (NPT) using the Nos\'{e}-Hoover thermostat with
temperature $T_0 = 0.6$ and pressure $P_0 =0.025$, a modified
velocity-Verlet integration scheme, and time step $\Delta {\overline
  t} = 10^{-3}$. We then cool the system into a glassy state at zero
temperature using a linear cooling ramp with time ${\overline t}$,
such that $T({\overline t}) = T_0-R {\overline t}$. (The cooling rate
is measured in units of $\epsilon_{AA}^{3/2}/(m^{1/2} \sigma_{AA})$,
where the Boltzmann constant $k_B=1$.)  We varied the cooling rate $R$
over more than three orders of magnitude, yet we ensured that $R$ was
larger than the critical cooling rate $R_c$ to avoid
crystallization. We vary $N$ from $250$ to $8000$ atoms to assess the
finite size effects.

% Excitation of the fundamental mode.
After cooling the system to zero temperature, we remove the periodic
boundary conditions in the $x$-, $y$-, and $z$-directions (creating
free surfaces) and then apply conjugate gradient energy minimization
to yield the zero-temperature configuration of the resonator,
$\textbf{R}_0=\{x_1,y_1,z_1,...,x_N,y_N,z_N\}$.  (See
Fig.~\ref{fig:method} (a).) To induce vibrations, we excite the
fundamental mode, i.e. the lowest eigenfrequency $\omega_1$ of the
dynamical matrix~\cite{tanguy2002continuum}, evaluated at
$\textbf{R}_0$. (See Fig.~\ref{fig:method} (b).)  The elongated, thin
shape of the resonator guarantees that the lowest eigenfrequency is
well-separated from higher ones. We then set the initial velocities of
the atoms, such that $\textbf{v} =
\{v_{x1},v_{y1},v_{z1},...,v_{xN},v_{yN},v_{zN}\} = \delta
\textbf{e}_1$, where $\textbf{e}_1$ is the eigenvector corresponding
to $\omega_1$ and $\delta = \sqrt{2NK/m}$, and run MD simulations at
constant total energy for a given time $t = \omega_1 {\overline
  t}/2\pi$. (The eigenvectors are normalized such that $\textbf{e}_i
\cdot \textbf{e}_j = \delta_{ij}$, where $\delta_{ij}$ is the
Kronecker delta and $i$,$j=1,\ldots,3N-6$ are the indexes of the 
eigenvectors that correspond to the non-trivial eigenfrequencies.)

\begin{figure}
\begin{center}
\includegraphics[width=0.9\columnwidth]{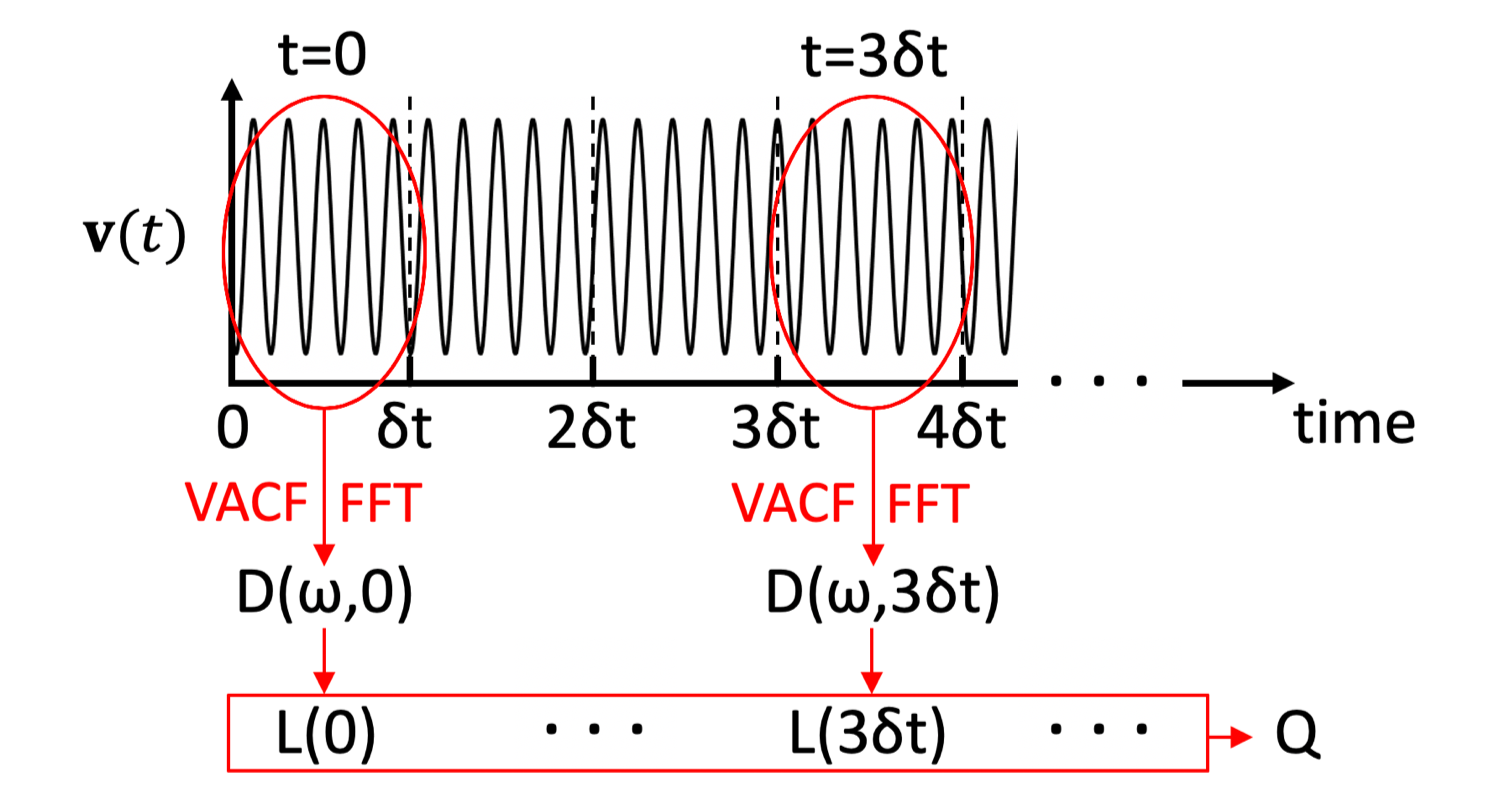}
\caption{Schematic diagram that illustrates the method we use to calculate the 
loss $L$ and quality factor $Q$ of model metallic glass resonators. 
We track the velocities $\textbf{v}$ 
of the atoms in the resonator over a long time period. The time 
series is broken up into $20$ time intervals with equal duration  $\delta t$. 
We calculate the velocity auto-correlation function (VACF) for each time 
interval, 
and fast Fourier transform (FFT) it to measure the density of vibrational 
modes $D(\omega,t)$ and loss $L(t)$ for each time interval $t$. 
Using Eq.~\ref{q}, we can calculate the quality factor $Q$ from $L(t)$.}
\label{fig:timeslot}
\end{center}
\end{figure}

We track the atom positions and velocities over long time periods $t >
2700$ during the MD simulations. We then divide the long time series
into $20$ time intervals with equal duration $\delta t = 135$.  We
characterize the vibrational response of the system using two
methods. In the first, we determine the vibrational response using the
time period from $0$ to $\delta t$.  For the second method, we
quantify how the vibrational response varies in time following the
initial perturbation using a fixed tape length $\delta t$ for each
time interval. (We show that our results do not depend strongly on
tape length $\delta t$ in Appendix~\ref{sec:appendix_tapelength}.)

For each time interval between $t$ to $t+\delta t$, we calculate
the Fourier transform of the velocity autocorrelation function to
determine the density of vibrational modes $D(\omega,
t)$~\cite{bertrand2014hypocoordinated}:
\begin{equation}
\label{loss}
D(\omega, t)=\int_0^{\delta t}{\langle \textbf{v}(t_0+\tau) \cdot \textbf{v}(t_0) \rangle_{t} e^{i\omega\tau}d\tau},
\end{equation}
where $\omega$ is the angular frequency and $\langle.\rangle_{t}$ indicates 
an average over all atoms and time origins $t_0$ between $t$ and 
$t+\delta t$.  See Fig.~\ref{fig:timeslot} for a summary of this approach. 

\begin{figure}
\begin{center}
\includegraphics[width=0.95\columnwidth]{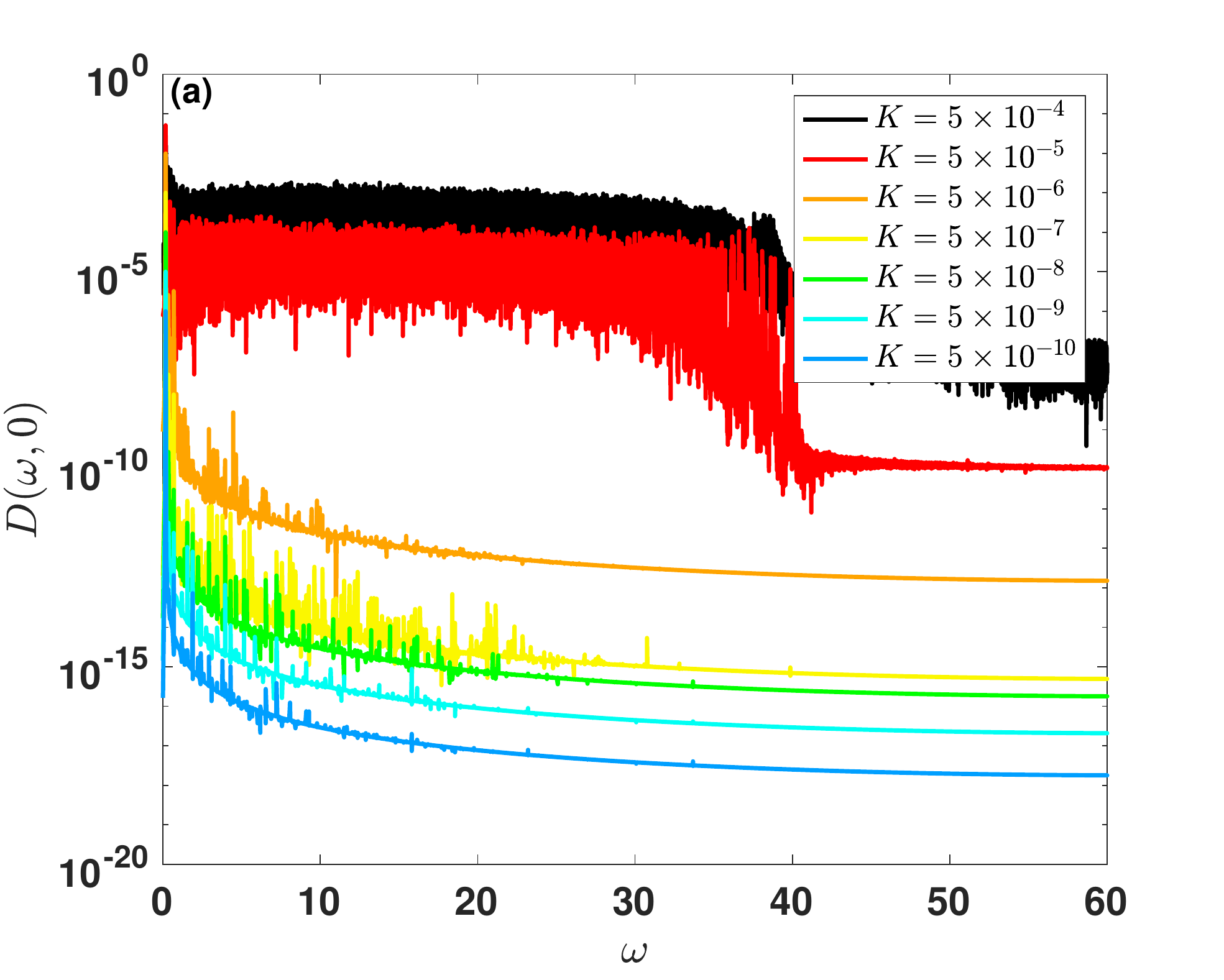}
\includegraphics[width=0.95\columnwidth]{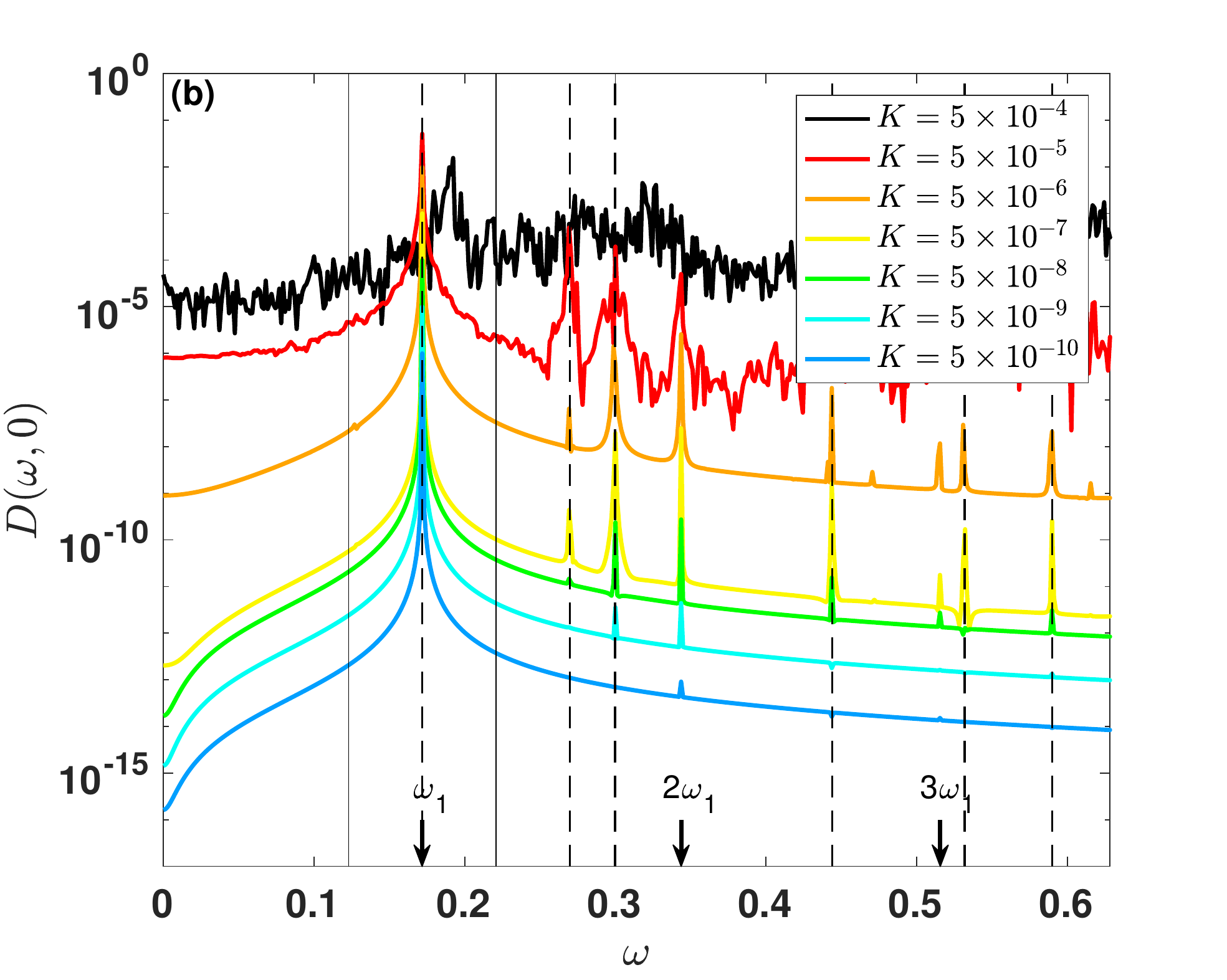}
\caption{(a) The density of vibrational modes $D(\omega,0)$ for the time 
interval $t=0$ as a function of the kinetic energy per atom $K$.
(b) $D(\omega,0)$ for the same systems in (a), but a close-up of the low 
frequency regime. The vertical dashed lines indicate 
the vibrational frequencies ($\omega_1,\omega_2,\ldots,\omega_{3N-6}$) 
calculated from the dynamical matrix. 
The arrows indicate integer multiples of the fundamental frequency 
$\omega_1$ and the two vertical solid lines show the region of 
frequencies near $\omega_1$ used in Eq.~\ref{loss} to calculate 
the loss.}
\label{fig:spec}
\end{center}
\end{figure}

% Loss and Q-factor analysis.
For each time interval, we also determine the fraction of the kinetic 
energy that has transferred from the fundamental mode (with frequency $\omega_1$) to other frequencies by defining
the loss,
\begin{equation}
\label{loss}
L(t)=1-\frac{\int_{\omega_1-\Delta\omega}^{\omega_1+\Delta\omega} D(\omega,t)d\omega}{\int_0^{\infty} D(\omega,t)d\omega}.
\end{equation}
where $\Delta\omega=(\omega_2-\omega_1)/2$. (See Fig.~\ref{fig:spec}
(b).) By determining the loss $L(t)$ over consecutive time intervals,
we can calculate the quality factor
\begin{equation}
\label{q}
Q=\omega_1\left(\frac{dL(t)}{dt}\right)^{-1}.
\end{equation}
Note that the results do not depend strongly on the magnitude of
$\Delta \omega$ as long as it brackets $\omega_1$. 

To track the atomic displacements during vibration, we will also calculate
the root-mean-square deviation (RMSD) between two  
configurations, e.g. $\textbf{R}(t_1)$ and $\textbf{R}(t_2)$ at different 
times $t_1$ and $t_2$:
\begin{widetext}
\begin{equation}
\label{rmsd}
d(\textbf{R}(t_1),\textbf{R}(t_2))=\sqrt{ N^{-1} \sum_{i=1}^{N}{(x_{i}(t_1)-
x_{i}(t_2))^2+(y_{i}(t_1)-y_{i}(t_2))^2+(z_{i}(t_1)-z_{i}(t_2))^2}}, 
\end{equation}
\end{widetext}
where the sum is over all atoms. 

\section{Results} 
\label{sec:results}

The results are organized into three sections. In
Sec.~\ref{sec:results_dissipation}, we quantify the density of
vibrational modes $D(\omega,0)$ and loss $L(0)$ during the first time
interval ($t=0$) as a function of the initial kinetic energy per atom
$K$, and investigate the effects of atomic rearrangements on the
vibrational response. We also study the dependence of $D(\omega,t)$
and $L(t)$ on the time interval $t$ and calculate the quality factor
$Q$. We identify three characteristic regimes for vibrational response
as a function of $K$: the linear response regime, where there is no
leakage of energy from the fundamental mode to others, the nonlinear
regime, where energy leakage occurs at short times, but it stops at
long times, and the strong loss regime where atomic rearrangements
occur, causing large losses and small $Q$. In
Sec.~\ref{sec:results_improvement}, we investigate how variations of
the system size $N$ and cooling rate $R$ affect the frequency of
atomic rearrangements, and thus the vibrational response.  In
Sec.~\ref{sec:results_X}, we calculate the loss in resonators made
from polycrystalline and defected crystalline materials and compare it
to resonators made from amorphous materials. We find that the losses
generated from resonators with amorphous structure are comparable to
that for crystalline resonators, and thus glassy materials may be
promising for high-$Q$ resonator applications.

\subsection{Intrinsic dissipation: Nonlinearity and Atomic Rearrangements}
\label{sec:results_dissipation}

We first focus on model metallic glass resonators with $N=2000$
generated using cooling rate $R=10^{-2}$. In Fig.~\ref{fig:spec}, we
show the density of vibrational modes $D(\omega,0)$ after exciting the
system along the fundamental mode as a function of the kinetic energy
per atom over six orders of magnitude from $K=5\times10^{-10}$ to
$5\times10^{-4}$. When $K$ is small, most of the response remains in
the fundamental mode, $\omega_1$, indicating that the system is in the
linear response regime. As $K$ increases, energy begins to leak to
other modes of the dynamical matrix (indicated by the dashed vertical
lines in Fig.~\ref{fig:spec} (b)), as well as harmonics of the
fundamental mode (indicated by the arrows in Fig.~\ref{fig:spec}
(b)). The leakage of energy from the fundamental mode is due to the
nonlinearity of the Lennard Jones potential near the minimum and not
due to the cutoff at $r_c = 2.5 \sigma_{ij}$~\cite{mizuno2016cutoff}.
To test this, we also carried out studies of weakly nonlinear springs
with $r_c \rightarrow \infty$ and found similar results. 

\begin{figure}
\begin{center}
\includegraphics[width=0.49\columnwidth]{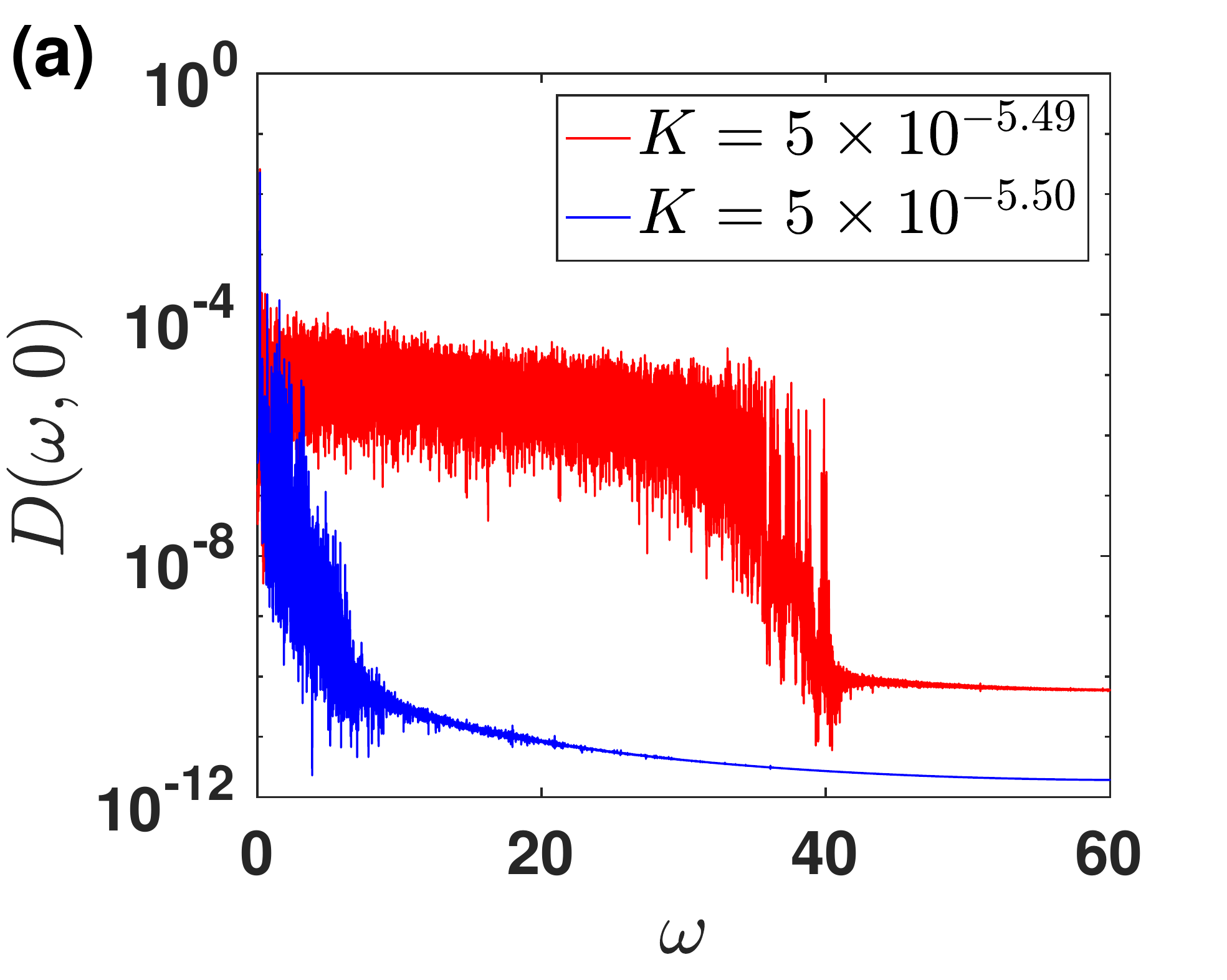}
\includegraphics[width=0.48\columnwidth]{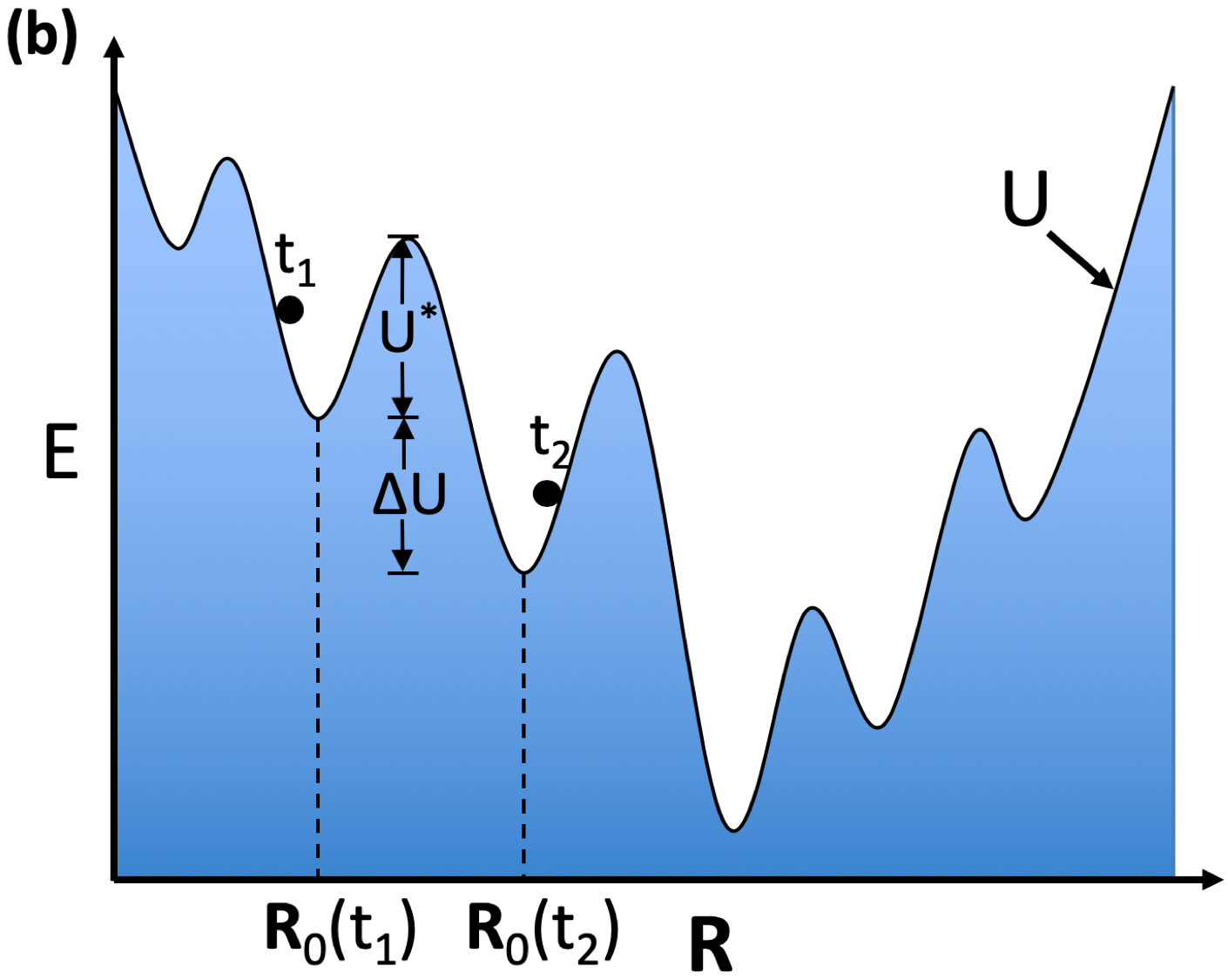}
\includegraphics[width=0.95\columnwidth]{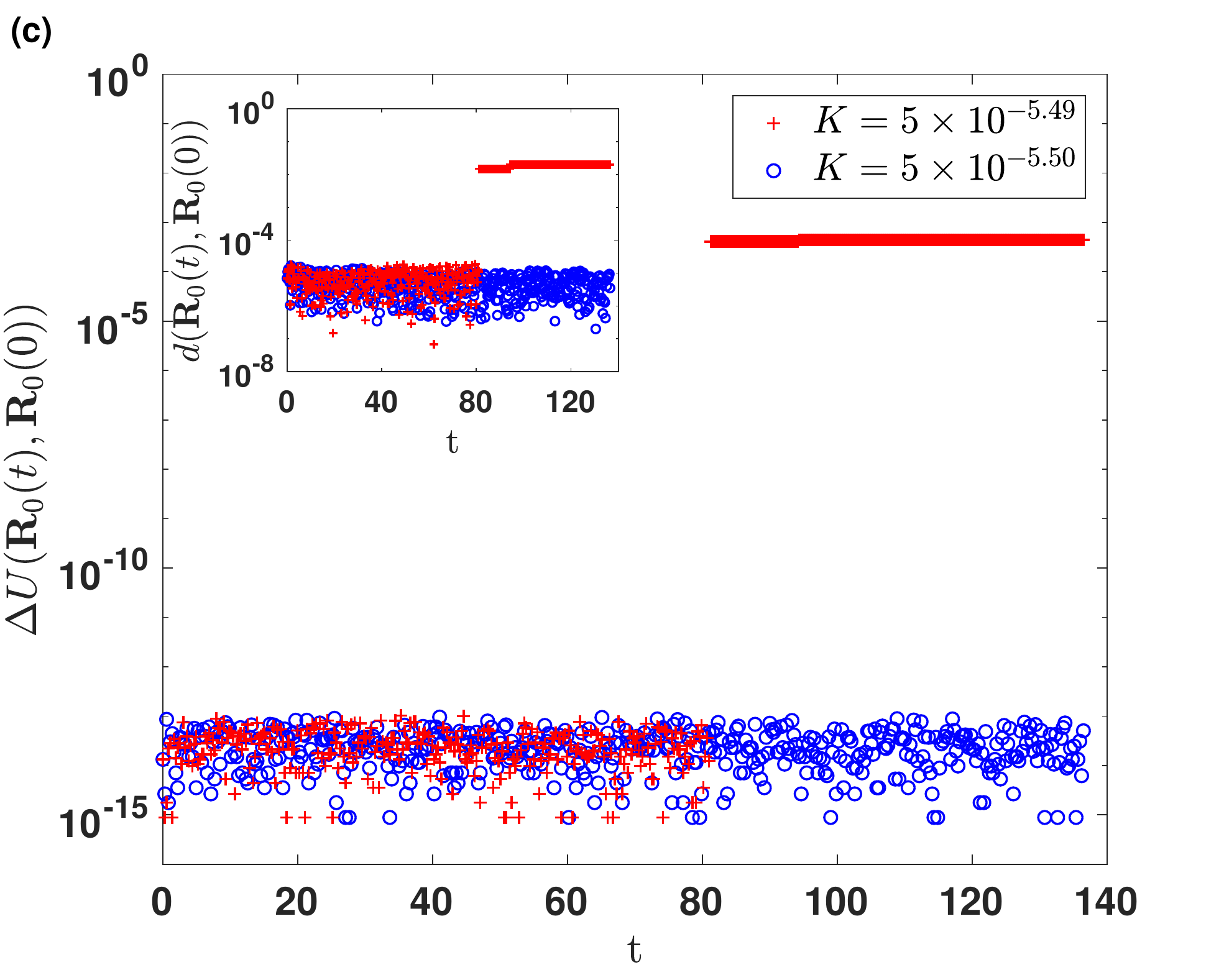}
\caption{(a) Density of vibrational modes $D(\omega,0)$ for 
$K=5\times10^{-5.50}$ (blue) and $5\times10^{-5.49}$ (red). 
(b) Schematic diagram of the energy landscape with  
axes, the total energy per atom $E=U+K$ and atomic configuration 
$\textbf{R}$. The configurations $\textbf R_0(t_1)$ and 
$\textbf R_0(t_2)$ represent the inherent structures ({\it i.e.}, the nearest local 
potential energy minima) of the vibrating system
at times $t_1$ and $t_2$, respectively.  $\Delta U$ is the difference 
in the potential energy per atom and $U^*$ is the energy barrier 
between the configurations $\textbf R_0(t_1)$
and $\textbf R_0(t_2)$. (c) $\Delta U(\textbf{R}_0(t),\textbf{R}_0(0))$ 
between the inherent structures at times $t$ and $0$ for 
$K=5\times10^{-5.50}$ (blue circles) and $5\times10^{-5.49}$ (red pluses).
The inset shows the root-mean-square deviation (RMSD) $d(\textbf{R}_0(t),\textbf{R}_0(0))$ between the inherent 
structures at times $t$ and $0$.}
\label{fig:rearrange}
\end{center}
\end{figure}

In Fig.~\ref{fig:spec}, we show that there is a qualitative change in
the vibrational response when $K$ increases from $5\times10^{-6}$ to
$5\times10^{-5}$.  At the higher value of $K$, the vibrational
response is noisy and energy is redistributed over a much wider range
of frequencies than at the lower value of $K$.  A more refined search
shows that this qualitative change occurs in the kinetic energy
interval $5 \times 10^{-5.50} < K_r < 5 \times 10^{-5.49}$ as shown in
Fig.~\ref{fig:rearrange} (a).

We now investigate the cause for the qualitative change in the
vibrational response for $K > K_r$. To do this, for each fluctuating
configuration $\textbf{R}(t)$, we calculate the corresponding
inherent structure, or the configuration of the nearest local potential 
minimum $\textbf{R}_0(t)$,
using conjugate gradient energy minimization. A schematic illustrating the
potential energy landscape is shown in Fig.~\ref{fig:rearrange}
(b). In Fig.~\ref{fig:rearrange} (c), we plot the difference in the
potential energy per atom $\Delta U(\textbf{R}_0(t),\textbf{R}_0(0))
=|U(\textbf R_0(t))-U(\textbf R_0(0))|$ as a function of time for $K <
K_r$ and $K>K_r$.  When $K < K_r$, $\Delta U \sim 10^{-14}$ for all
times, indicating that the system remains in the basin of the inherent
structure at $t=0$. For $K > K_r$, $\Delta U$ jumps from $\sim
10^{-14}$ to $\sim 10^{-3}$ near $t^* \sim 80$, indicating that the
system transitions from the basin of the inherent structure at $t=0$
to that of a different inherent structure at $t^*$ following an atomic
rearrangement. We also used Eq.~\ref{rmsd} to calculate the root-mean-square 
deviation
between the inherent structures $\textbf{R}_0(0)$ and 
$\textbf{R}_0(t)$ at times $0$ and $t$ during the vibrations.  
In the inset of
Fig~\ref{fig:rearrange} (c), we show that $\Delta U$ and $d$ display 
similar behavior. For $K < K_r$, $d \sim 10^{-6}$ for all times.  For 
$K > K_r$, near $t^* \sim 80$, $d$ jumps from $\sim 10^{-6}$ to $\sim 
10^{-2}$ again indicating that an atomic rearrangement occurs at $t^*$.
One can also see that subsequent rearrangements occur at later times, 
which are indicated by jumps in $\Delta U$ and $d$.  
These results emphasize that atomic rearrangements induce significant 
redistribution of energy from the fundamental mode to other frequencies.  

\begin{figure}
\begin{center}
\includegraphics[width=0.95\columnwidth]{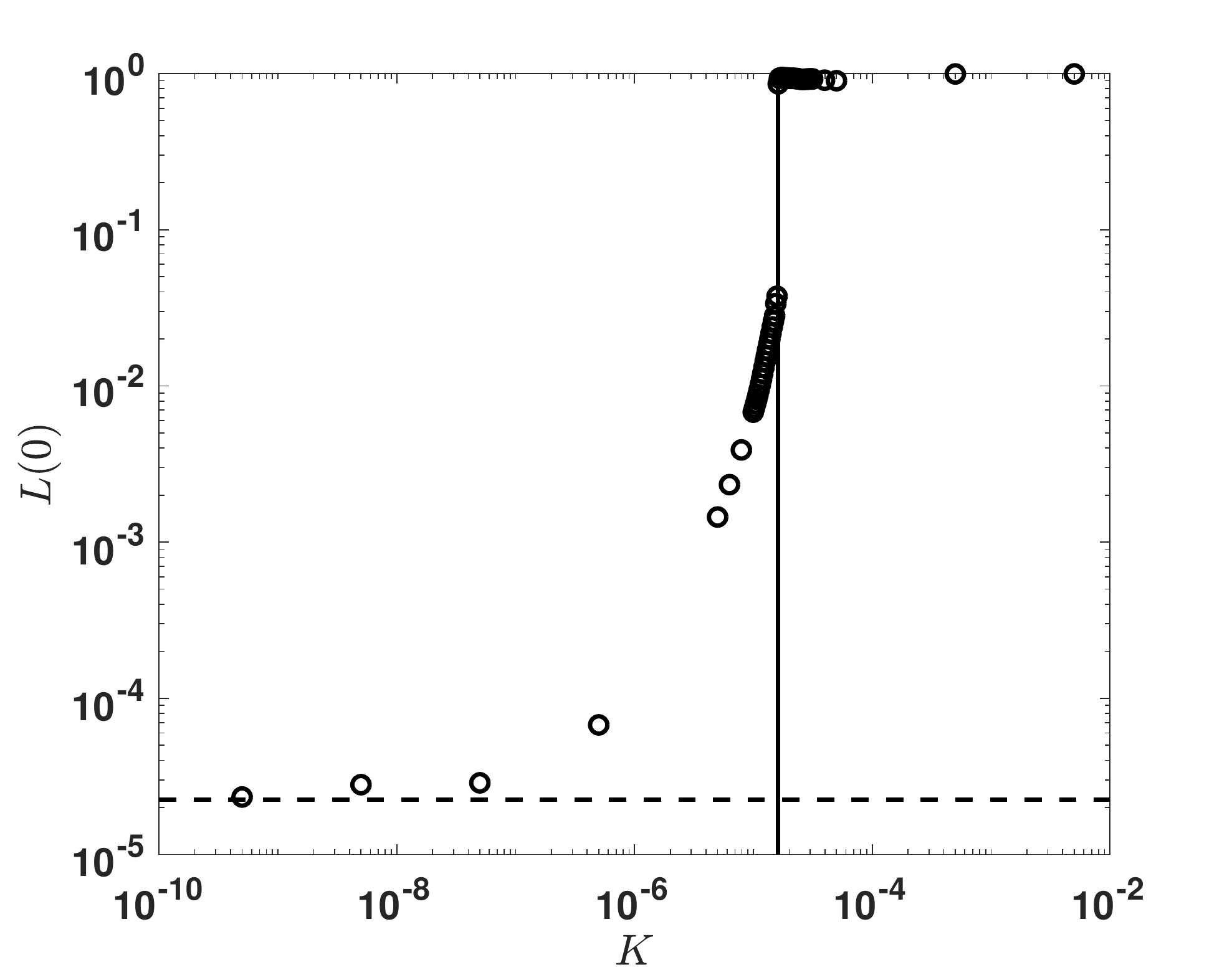}
\caption{Loss $L(0)$ (Eq.~\ref{loss}) for the first time interval $t=0$ 
versus the 
initial kinetic energy per atom $K$.
The solid vertical line indicates $K_r \approx 5\times10^{-5.49}$ at which 
the first atomic rearrangement occurs.  
The dashed horizontal line indicates the loss threshold $L_{l}$ for 
a harmonic oscillator with a measurement time $\delta t$ that 
deviates from an integer.}
\label{fig:loss}
\end{center}
\end{figure}

We quantify the leakage of energy from the fundamental mode to other
frequencies over the first time interval $t=0$ by calculating the loss
$L(0)$ (defined in Eq.~\ref{loss}) as a function of $K$ in
Fig.~\ref{fig:loss}.  We calibrate the measurement of the loss by
studying perfect cosine oscillations of the velocity of a single atom
over a tape length of $\delta t$.  Since in general $\delta t$ is not
an exact integer multiple of the oscillation period, the loss $L_l
\sim 10^{-4.5}$ we measured for a cosine wave is small, but
nonzero. We find that the lower threshold for the loss $L_l$ does not
affect the results we present. In Fig.~\ref{fig:loss}, we show that at small $K$,
$L(0) \sim L_l$ and $L(0)$ increases smoothly with increasing $K$
until reaching $0.04$ near $K_r$. At $K_r$, the loss jumps to $L(0) \sim 1$,
indicating the onset of atomic rearrangements, and remains there for $K
> K_r$.

\begin{figure}
\begin{center}
\includegraphics[width=0.95\columnwidth]{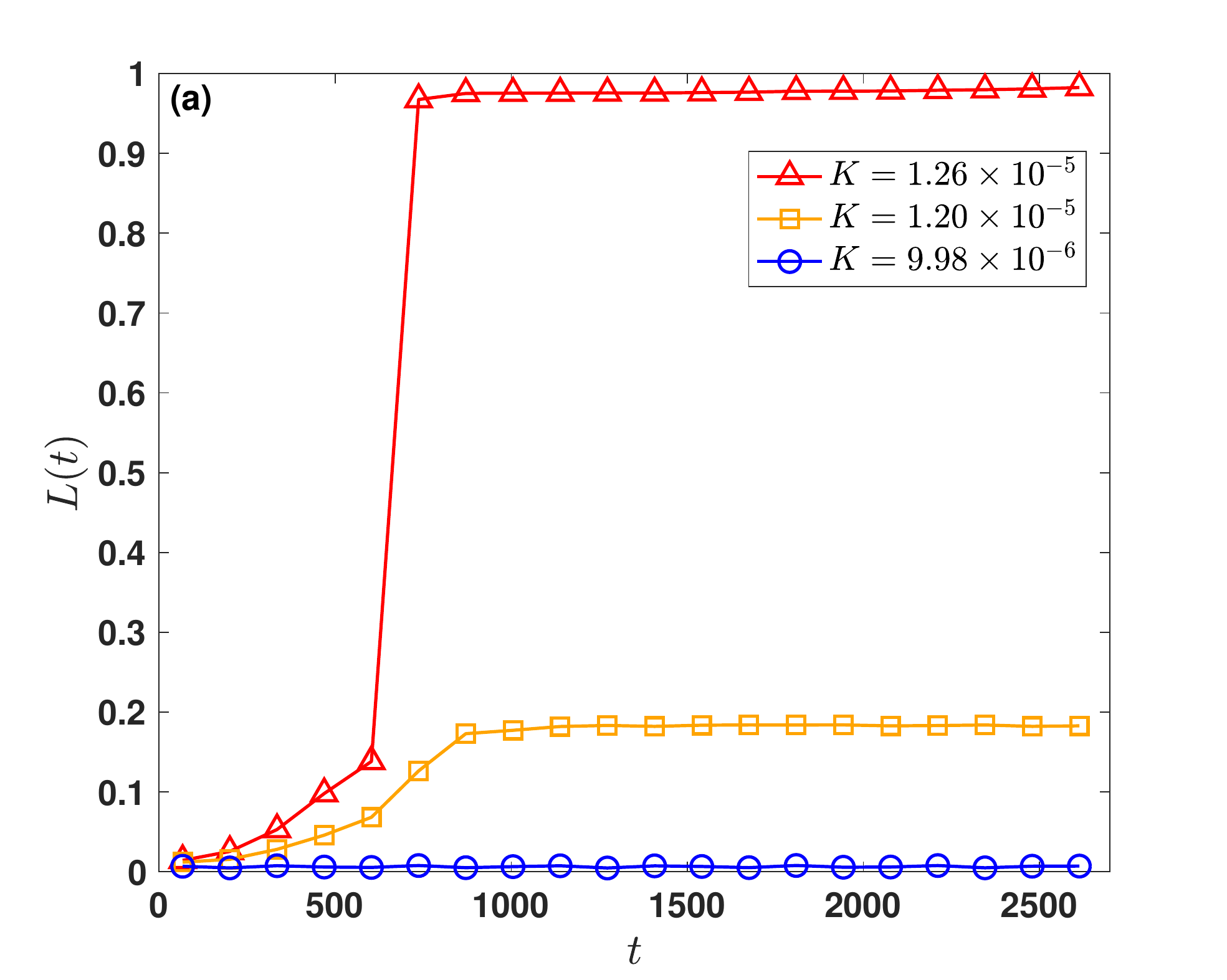}
\includegraphics[width=0.95\columnwidth]{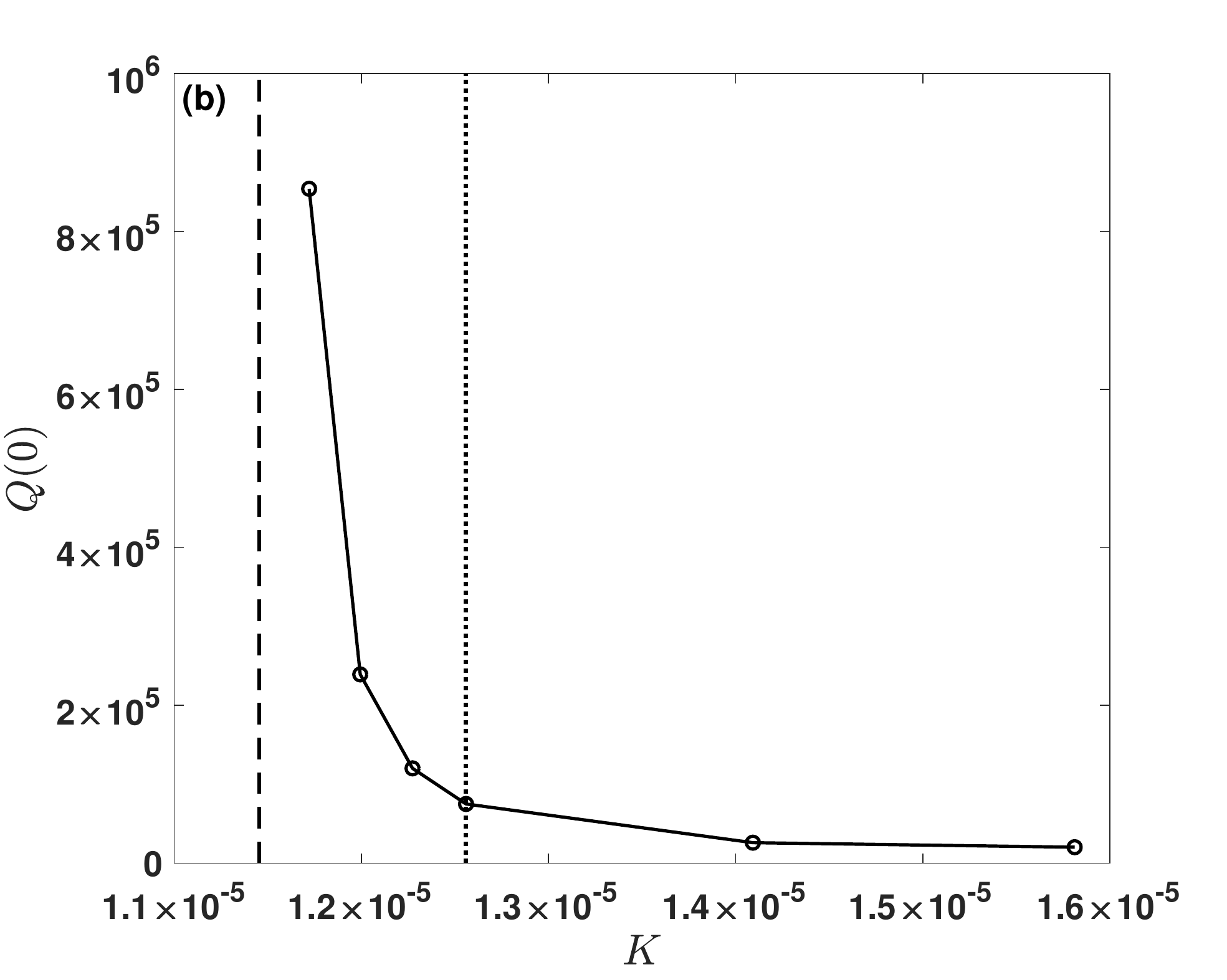}
\caption{(a) Loss $L(t)$ versus the time interval $t$ for kinetic energy
per atom 
$K=9.98\times 10^{-6}$ (blue circles), $1.20\times 10^{-5}$ (orange squares), and 
$1.26\times 10^{-5}$ (red triangles). (b) Quality factor $Q(0)$ for the 
first time interval $t=0$ as a function of $K$. 
The dashed vertical line indicates $K_{nl} \approx 1.15\times 10^{-5}$ 
at which $Q(0) \rightarrow \infty$. The vertical dotted line indicates 
$K_r \approx 1.26\times 10^{-5}$, above which atomic rearrangements occur.}
\label{fig:q}
\end{center}
\end{figure}

In Fig.~\ref{fig:loss}, we showed the loss for only the first time
interval $t=0$.  We characterize the time-dependent loss in
Fig.~\ref{fig:q} (a). 
(We also include the variation of the density of states $D(\omega,t)$ with time $t$
in Appendix~\ref{sec:appendix_dissipating}.)
We identify three distinct regimes. First, when
$K < K_{nl}$, with $K_{nl} \approx 1.15 \times 10^{-5}$ the loss
$L(t)$ is small and does not increase with $t$, and thus $Q
\rightarrow \infty$. In the second regime, for intermediate $K_{nl} <
K < K_r$ (such as $K=1.20\times 10^{-5}$ in Fig.~\ref{fig:q} (a)),
$L(t)$ initially increases with $t$ smoothly, generating a finite $Q$,
but then $L(t)$ reaches a plateau and $Q \rightarrow \infty$ at long
times. In the third regime, for $K>K_r$ (such as $K=1.26\times
10^{-5}$ in Fig.~\ref{fig:q} (a)), $L(t)$ increases with $t$ smoothly
(indicating finite $Q$), until an atomic rearrangement event occurs
and $L(t)$ jumps to a large value $L\sim 1$. $L(t)$ continues to increase 
after the first atomic rearrangement. 

We evaluate $Q(0)$ for the first time interval $t=0$ using
Eq.~\ref{q}, and show the results as a function of $K$ in
Fig~\ref{fig:q} (b).  We find that $Q(0) \sim 2 \times 10^4$ for
$K\sim 1.5\times 10^{-5}$ and $Q(0)$ increases with decreasing
$K$. For $K \lesssim K_{r}$, $Q$ begins to increase sharply, diverging
as $K \rightarrow K_{nl}$, indicating the behavior for a perfect
linear resonator for $K < K_{nl}$.  These results indicate that to
design a high-$Q$ metallic glass resonator, one needs to fabricate a
system with a large value for $K_r$ and operate the resonator at $K < K_r$.

\begin{figure}
\begin{center}
\includegraphics[width=0.92\columnwidth]{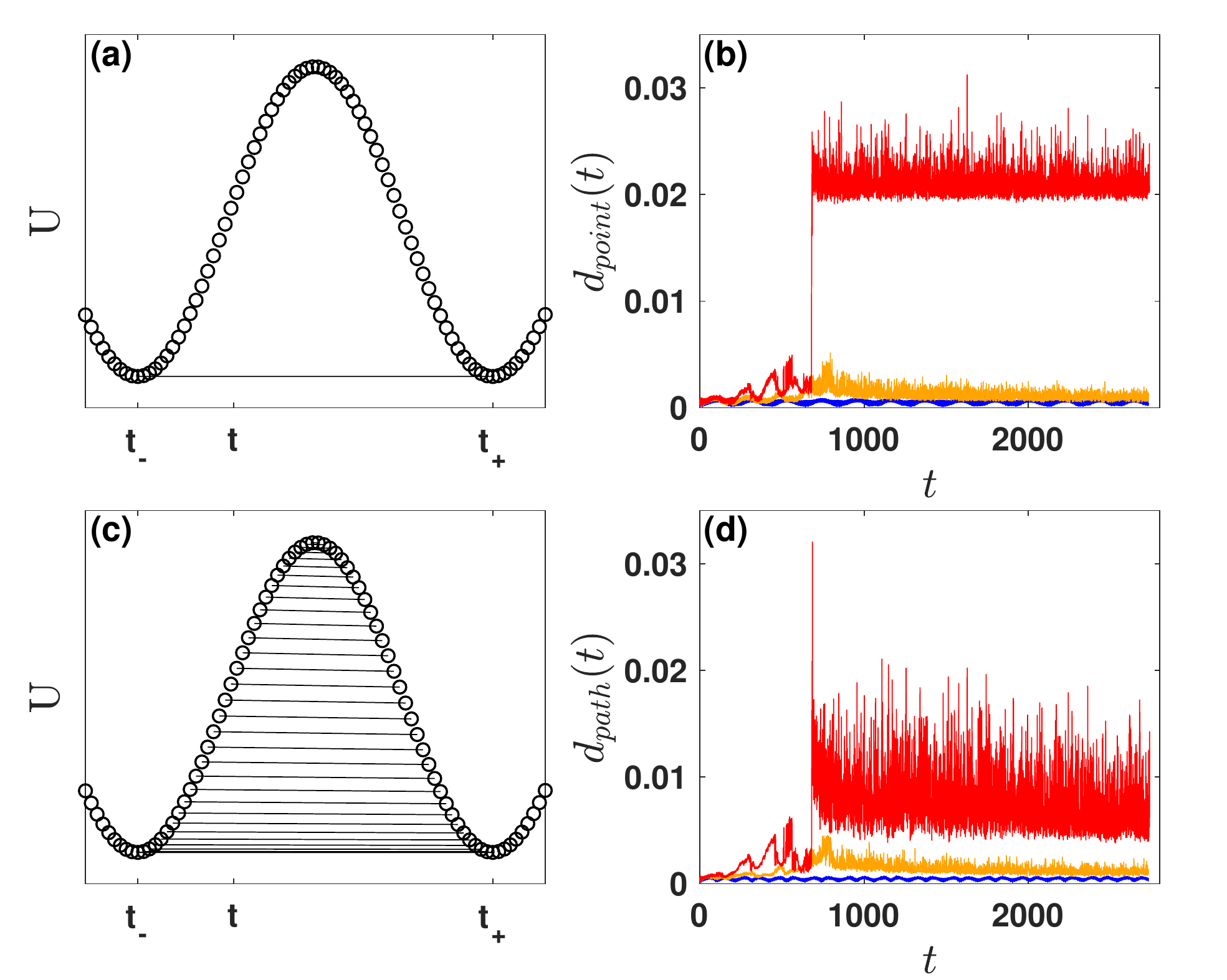}
\includegraphics[width=0.9\columnwidth]{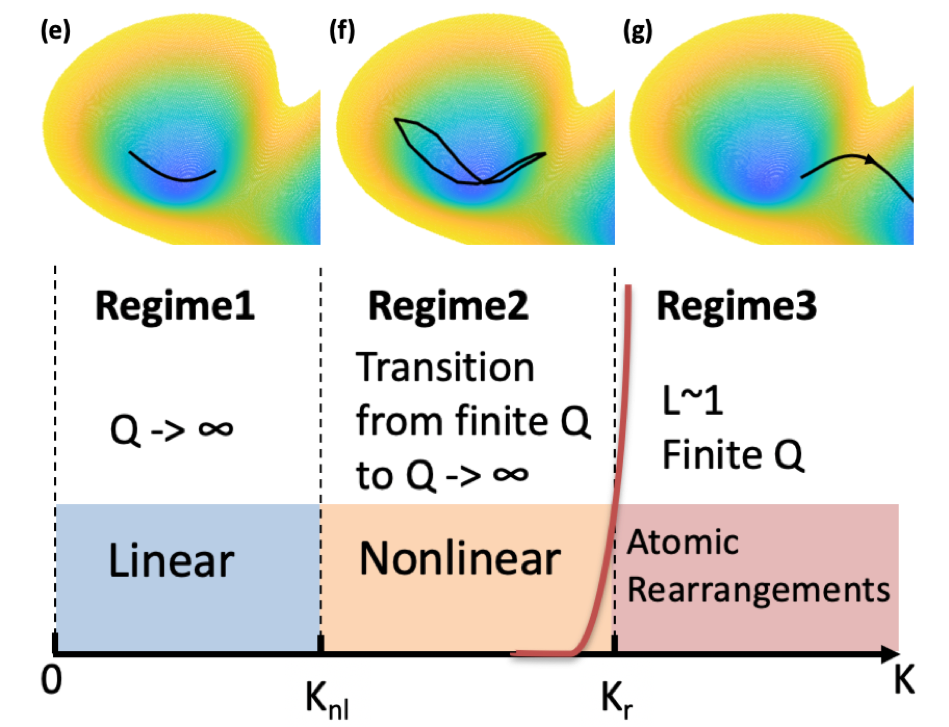}
\caption{(a) and (c) Total potential energy per atom $U$ as a function of 
time from $0$ to $n\delta t$, where $n=20$. The times $t_-$ and $t_+$ indicate 
successive times at which $U$ is at a minimum during each half cycle. The horizontal solid line in 
(a), connecting the times $t_-$ and $t_+$, indicates the times for each 
half cycle at which the 
RMSD, $d_{\rm point}(\textbf{R}(t_-),\textbf{R}(t_+))$, is calculated in 
panel (b). In (b), we show $d_{\rm point}$ for $K=9.98\times 10^{-6}$ 
in the regime $K < K_{nl}$ (blue), $K=1.20\times 10^{-5}$ in the 
regime $K_{nl} < K < K_r$ (orange), and $K=1.26\times 10^{-5}$ in the regime 
$K > K_r$ (red).  The horizontal solid lines in (c) indicate the times $t_i$ 
and $t_j=t_++t_--t_i$ during
each half cycle that 
are used to calculate the RMSD, $d_{\rm path} =  
\langle d(\textbf R(t_i),\textbf R(t_j)) \rangle_{t_i}$, where the 
angle brackets indicate an average over the $1/2\Delta t \sim 300$ uniformly 
spaced times $t_i$. $d_{\rm path}$ in panel (d) is shown for the same 
values of $K$ as in (b). 
(e)-(g) Schematic diagram that shows the system trajectories (solid black lines)
in the potential energy landscape (shaded contours from high (orange) 
to low (blue) energies) for the three regimes of oscillations ($1$: $K < K_{nl}$, 
$2$: $K_{nl} < K < K_r$, and $3$: $K > K_r$). 
The red solid line indicates the probability of
an atomic rearrangement versus $K$.
}
\label{fig:regimes}
\end{center}
\end{figure}

To understand the nature of oscillations in metallic glass resonators
(e.g. $U(t)$ in Fig.~\ref{fig:regimes} (a) and (c)), we calculate the
point RMSD $d_{\rm point}(t)$ and path RMSD $d_{\rm path}(t)$ in
Fig.~\ref{fig:regimes} (b) and (d).  $d_{\rm point}(t)$ quantifies the
deviations in the configurations that are the closest to the potential
energy minimum in each half cycle, and $d_{\rm path}(t)$ quantifies the
deviations in the configurations at corresponding times before and
after the turning point of the oscillation during each half
cycle. 

When $K < K_{nl}$ (regime $1$), the system is in the linear response
regime, the path in configuration space followed by the resonator is
nearly parabolic as shown in Fig.~\ref{fig:regimes} (e), and both
$d_{\rm point}$ and $d_{\rm path} \sim 0$.  When the system enters the
nonlinear regime, $K_{nl} < K <K_r$ (regime $2$), $d_{\rm point}$ and
$d_{\rm path}$ (as well as the loss $L(t)$ in Fig.~\ref{fig:q} (a))
increase with $t$ until $t^* \approx 790$.  For $t >t^*$, $d_{\rm
  point}$, $d_{\rm path}$, and $L(t)$ reach plateaus and then remain
nearly constant in time. This behavior indicates that the resonator is
undergoing {\it nonlinear} oscillations, in which the system does not
retrace the same configurations above and below the turning point for
each half cycle, but the system is nearly reversible. (See
Fig.~\ref{fig:regimes} (f).) In the third regime $K > K_r$, the
probability for an atomic rearrangement increases strongly. In this
regime, the system can traverse the saddle points, enter the basins
corresponding to new potential energy minima, and is thus microscopically 
irreversible.  The three regimes describing resonator oscillations 
are summarized in
Fig~\ref{fig:regimes} (e)-(g).

\subsection{Methods to increase $K_r$ and enhance $Q$}
\label{sec:results_improvement}

\begin{figure}
\begin{center}
\includegraphics[width=0.95\columnwidth]{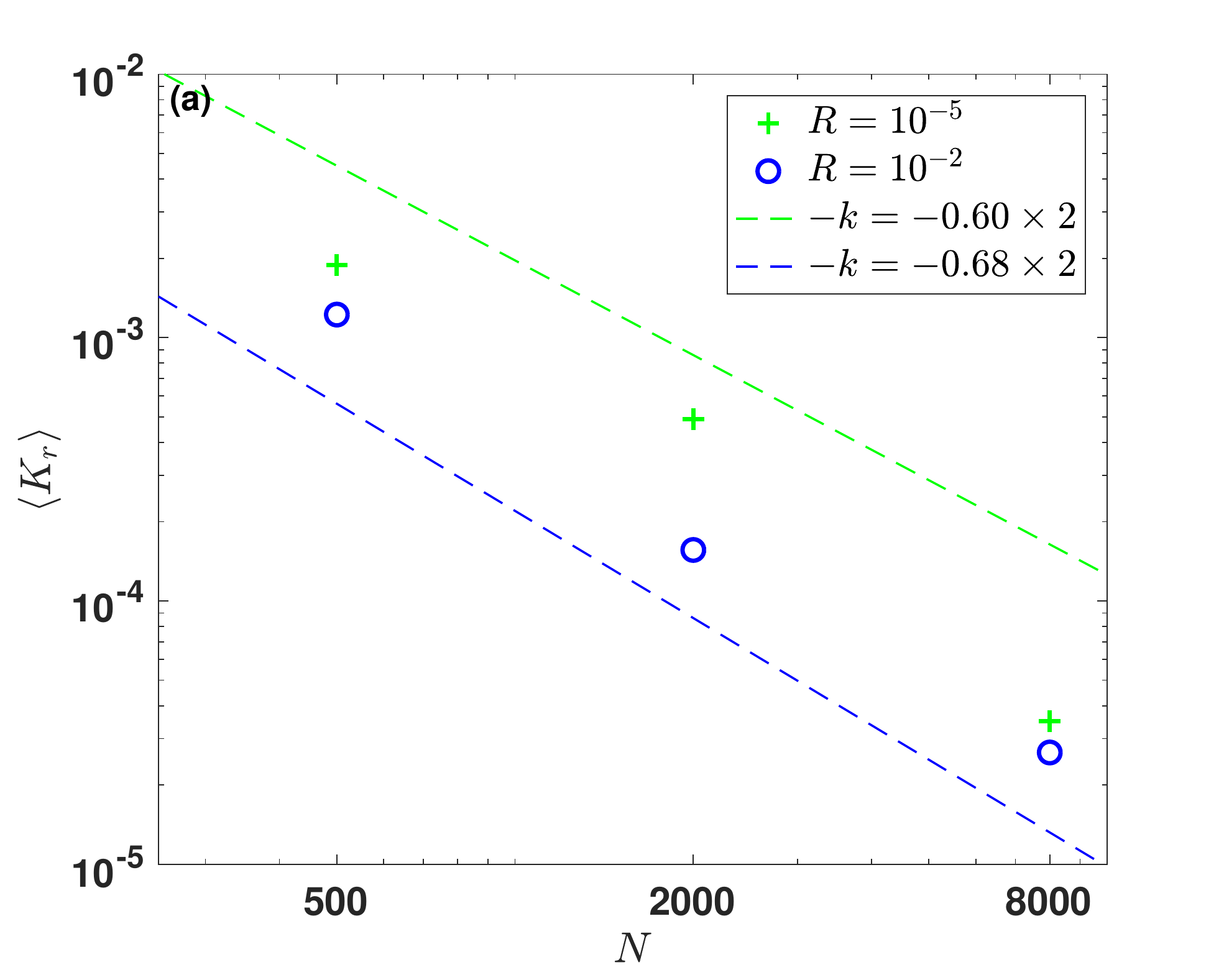}
\includegraphics[width=0.95\columnwidth]{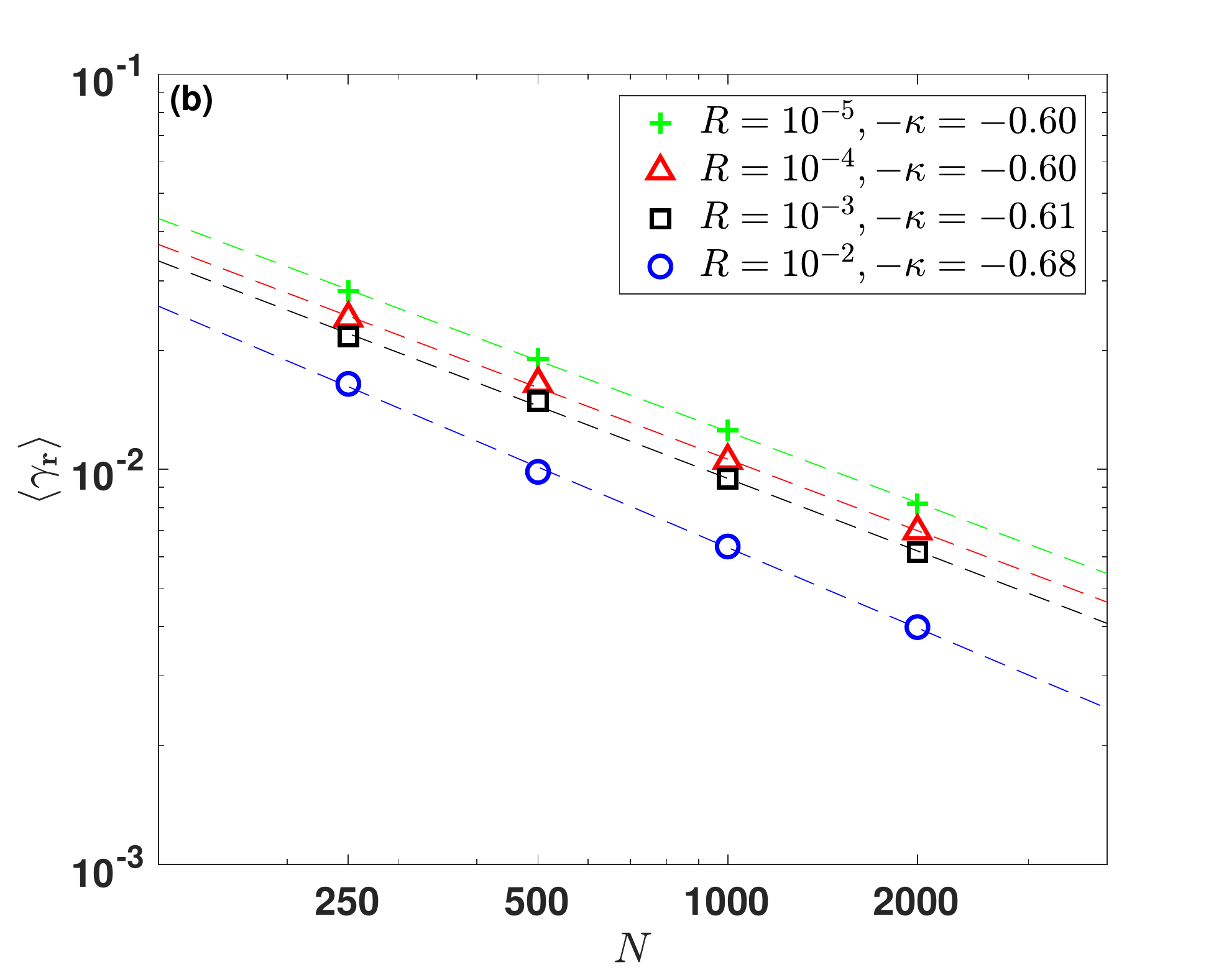}
\caption{(a) The ensemble-averaged kinetic energy per atom $\langle K_r \rangle$ above 
which the first atomic rearrangement occurs versus system size $N$ for 
rapidly and slowly cooled glasses with $R=10^{-2}$ (blue circles)
and $10^{-5}$ (green pluses), respectively. In each case, $\langle K_r \rangle$ 
is averaged over $20$ independent samples.  
The slopes of the dashed lines are $-k$.
(b) Ensemble-averaged pure shear strain $\langle \gamma_r \rangle$ above which the 
first atomic rearrangement occurs during athermal quasistatic pure shear 
versus system size $N$ for glasses prepared using $R=10^{-2}$ (blue circles), 
$10^{-3}$ (black squares), $10^{-4}$ (red triangles), and $10^{-5}$ 
(green pluses). $\langle \gamma_r \rangle$ is averaged over $500$ 
independent samples. The negative $\kappa$-values give the slopes of the dashed 
lines.}
\label{fig:scaling}
\end{center}
\end{figure}

In Sec.~\ref{sec:results_dissipation}, we showed that even single atomic
rearrangements give rise to significant loss and finite values of
$Q$. Thus, to generate high-$Q$ resonators, one must maximize $\langle
K_r \rangle \sim \langle U^* \rangle$, yielding systems with large
potential energy barriers.  In this section, we describe studies of
the ensemble-averaged $\langle K_r \rangle$ versus system size $N$ and
cooling rate $R$, averaged over typically $20$ independently generated
initial conditions. For each $R$ and $N$, we excite the resonator
along the fundamental mode corresponding to the lowest eigenvalue of
the dynamical matrix $\omega_1$ and monitor the system during the
first time interval $t=0$ as a function of $K$.

In Fig.~\ref{fig:scaling} (a), we show that the ensemble-averaged
kinetic energy per atom at which the first atomic rearrangement
occurs, $\langle K_r \rangle$, decreases with increasing $N$.  We find
that $\langle K_r \rangle \sim N^{-2 k}$, where $k \approx 0.68$ for
$R = 10^{-2}$ and $\approx 0.60$ for $R=10^{-5}$. $\langle K_r
\rangle$ is smaller for rapidly compared to slowly cooled glasses,
since $\langle U^* \rangle$ decreases with increasing
$R$~\cite{chen1971elastic,fan2014thermally,fan2017particle}. These results emphasize
that $Q$ can be increased by making resonators smaller and preparing
them using slower cooling rates.  For example, experimental studies of
Pt-based metallic glass micro-cantilevers have reported that the
quality factor can be increased by more than a factor of $3$ after
annealing~\cite{kanik2014high}.

We can also compare the kinetic energy per atom $\langle K_r \rangle$
required to induce the first atomic rearrangement in thermally vibrating
systems to the characteristic shear strain $\langle \gamma_r \rangle$
required to induce the first atomic rearrangement in systems driven by athermal
quasistatic (AQS) shear.  To calculate $\langle \gamma_r \rangle$, we
confine $N$ atoms interacting via the Kob-Andersen model to cubic
boxes with periodic boundary conditions in the $x$-, $y$-, and
$z$-directions.  We cool the samples from temperature $T_0$ to zero
using a linear ramp over a range of cooling rates $R$ from $10^{-5}$
to $10^{-2}$. For each sample, we perform AQS pure shear at fixed
volume $V$, i.e. at each strain step, we expand the box length and
move all atoms affinely in the $x$-direction by a small strain
increment $\delta\gamma_x=\delta\gamma=10^{-4}$ and compress the box
length and move all atoms affinely in the $y$-direction by the same
strain increment $\delta\gamma_y=-\delta\gamma$.  Following each
strain step, we perform conjugate gradient energy minimization at
fixed volume. To measure $\langle \gamma_r \rangle$, we employ the
method we developed previously~\cite{fan2017effects} to unambiguously
determine whether an atomic rearrangement occurs with an accuracy on
the order of numerical precision.  As shown in Fig.~\ref{fig:scaling}
(b), we find that $\langle \gamma_1 \rangle \sim N^{-\kappa}$ also
decreases with increasing $N$, where the system-size scaling
exponent $\kappa \sim 0.6$-$0.68$ is again only weakly dependent on
the cooling rate.  These results for the system-size scaling exponents 
in athermal quasistatic shear are consistent with dimensional arguments that suggest $\kappa \sim 2k$, and thus athermal quasistatic shear can be used to 
understand the low temperature properties of glasses.  

Using these results, we can estimate the strains below which
resonators can operate in the linear response regime. For the slowest
cooling rate $R=10^{-5}$, we find that $\log_{10} \langle \gamma_r
\rangle \approx -2k \log_{10} N + \gamma_{\infty}$, where $2k \approx
0.60$, $\gamma_{\infty} \approx 0.11$, $N = \rho(l/D)^3$, the number
density $\rho \approx 1.2$, $D \approx 3.7$~\AA~is a typical atomic
diameter for Ni$_{80}$P$_{20}$ metallic
glasses~\cite{sheng2012relating} (which is the subject of the
Kob-Andersen model), and $l$ is a characteristic lengthscale of the
resonator. We find that $\langle \gamma_r \rangle \sim 5 \times
10^{-4}$ for a resonator with $l \sim 20$~nm, whereas $\langle
\gamma_r \rangle \sim 3 \times 10^{-8}$ for a resonator with $l \sim
5$~ $\mu$m~\cite{poncharal1999electrostatic,purcell2002tuning}.
Micron-scale metallic glass resonators have been fabricated as
hemispherical shells~\cite{kanik2015metallic} and as
cantilevers~\cite{kanik2014high}. In addition, strains in the range
from $10^{-7}$ to $10^{-4}$ have been used in measurements of internal
friction in metallic glass resonators~\cite{khonik1996internal}.  Our
results emphasize that nano-sized metallic glass resonators operating
in the small strain regime regime (e.g. $< 10^{-7}$) are promising
high-$Q$ materials.
  
\subsection{Comparison between crystalline and amorphous resonators}
\label{sec:results_X}

In Sec.~\ref{sec:results_improvement}, we showed that the
characteristic kinetic energy per atom $\langle K_r \rangle$ above
which atomic rearrangements occur increases modestly with decreasing cooling
rate. Further, we know that crystalline ordering increases with
decreasing cooling rate. Does this imply that crystalline metals are
higher-$Q$ materials compared to amorphous metals?  In this section,
we calculate $\langle K_r \rangle$ for resonators made from single crystal,
polycrystalline, and defected crystalline materials and compare these
results to those for resonators made from homogeneously amorphous samples.

Crystalline metals often contain slip planes, dislocations, grain
boundaries, and other defects, and the defect density typically
increases with increasing cooling rate. To generate crystalline
materials with defects in simulations, we will again use the
Kob-Andersen model, but with monodisperse atoms, $\epsilon_{AA} =
\epsilon_{AB} = \epsilon_{BB} =1.0$ and $\sigma_{AA} = \sigma_{AB} =
\sigma_{BB} = 1.0$, to enhance crystallization.  We will employ the
same protocol as discussed in Sec.~\ref{sec:methods} to generate
thin-bar resonators with $N=2000$ and aspect ratio $L_x:L_y:L_z =
6:1:2$ over a range of cooling rates from $R=10^{-4}$ to
$10^{2}$. The method of excitation and measurement of the loss and
$K_r$ are also the same as described in Sec.~\ref{sec:methods}.

\begin{figure}
\begin{center}
\includegraphics[width=0.99\columnwidth]{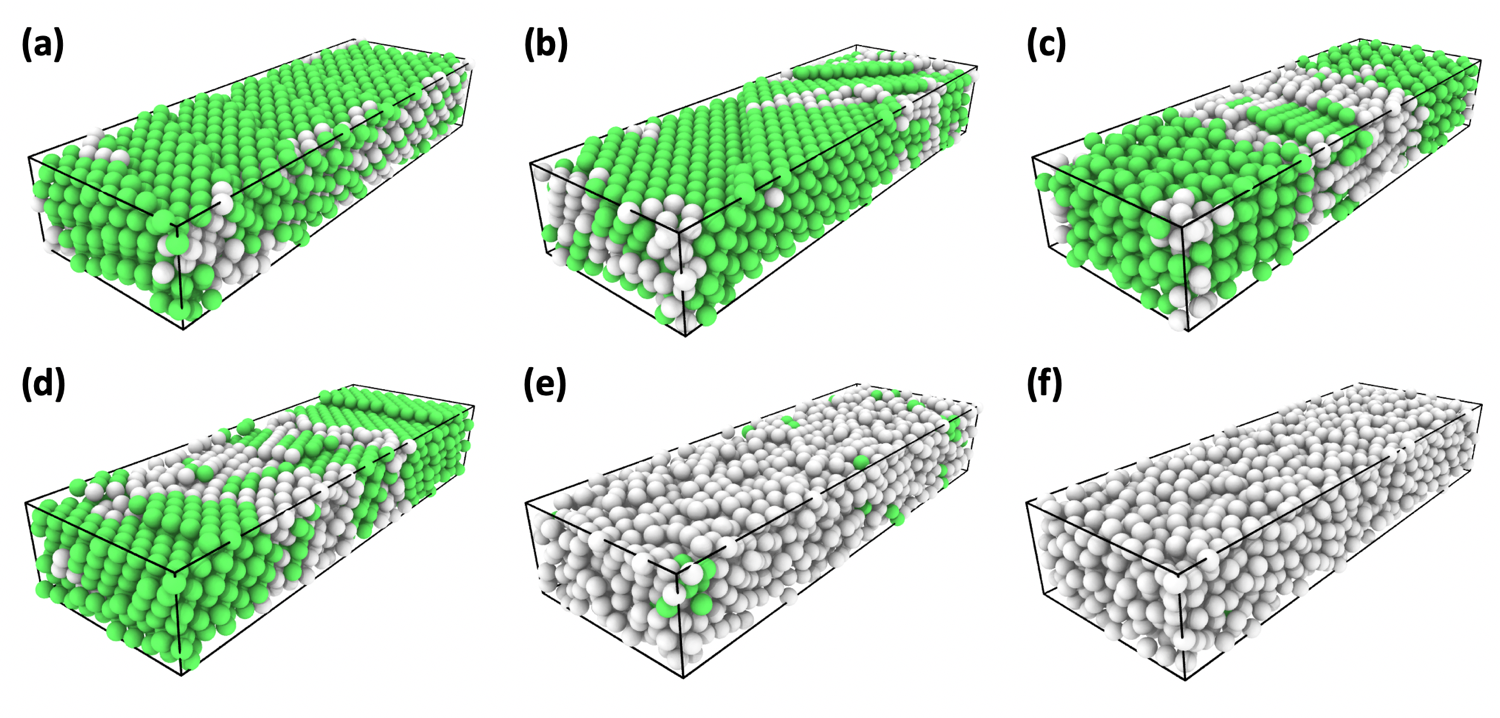}
\caption{Snapshots of thin-bar-shaped resonators with $N=2000$ monodisperse 
atoms obtained using cooling rates (a) $R=1.2 \times 10^{-4}$, (b) $1.2 \times 10^{-3}$, 
(c) $3 \times 10^{-3}$, (d) $6 \times 10^{-3}$, (e) $1.2 \times 10^{-2}$, and 
(f) $1.2 \times 10^{-1}$ in periodic boundary conditions prior to applying the excitations. Atoms with 
crystalline (FCC or HCP) order are colored green, while amorphous atoms are colored gray.}
\label{fig:Xdefects}
\end{center}
\end{figure}

\begin{figure}
\begin{center}
\includegraphics[width=0.96\columnwidth]{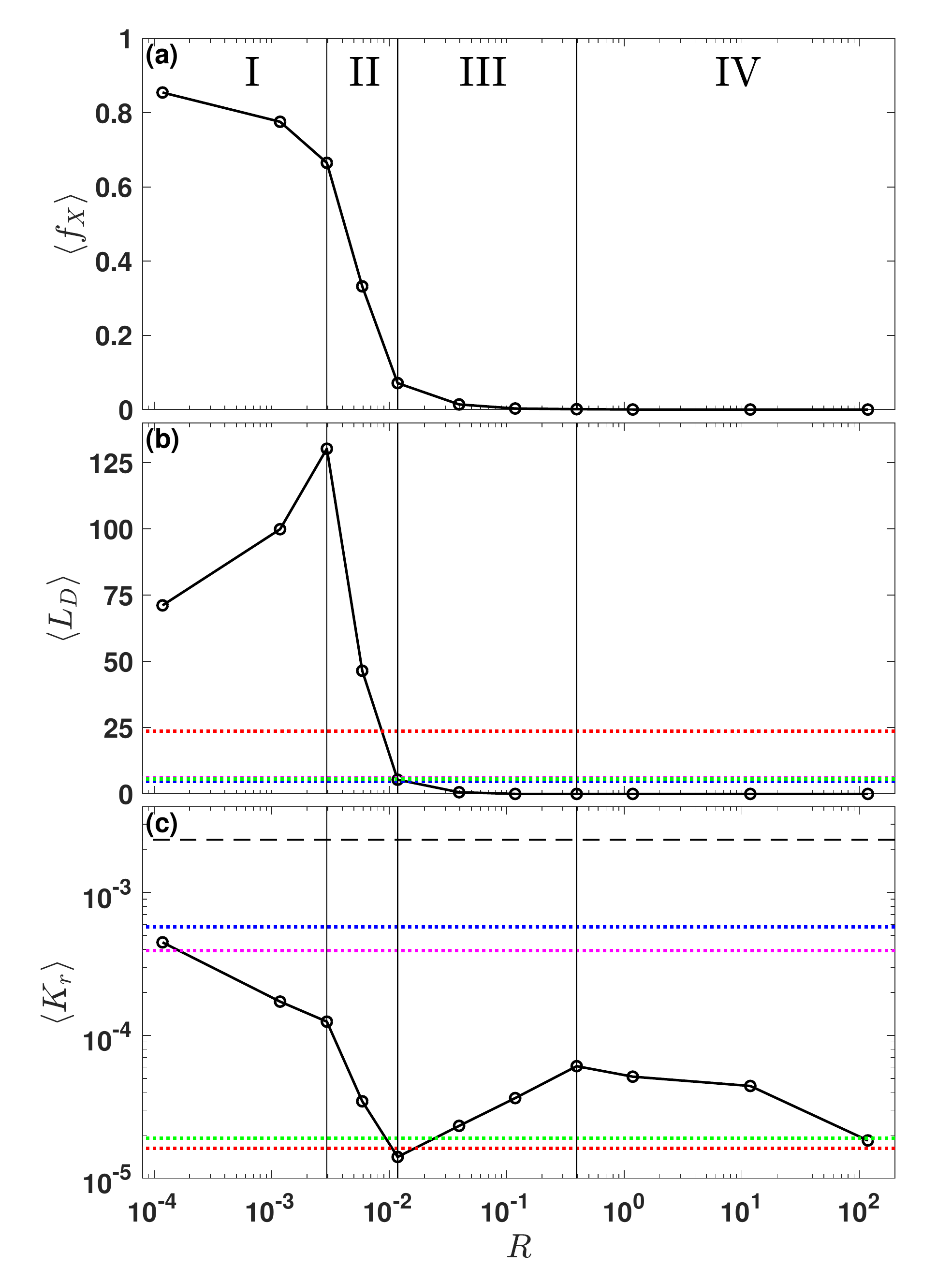}
\caption{(a) Fraction of crystalline atoms $\langle f_X \rangle$, 
(b) total dislocation length $\langle L_D \rangle$,
and (c) kinetic energy per atom $\langle K_r \rangle$ above which atomic 
rearrangements 
occur as a function of cooling rate $R$ for resonators made using the 
monodisperse Kob-Andersen model.  The ensemble averages are 
obtained by averaging over at least $10$ independent samples.
The dotted horizontal lines in (b) and (c) show $\langle L_D \rangle$ 
and $\langle K_r \rangle$ for four nearly perfect, crystalline thin bars with
specifically placed defects. For example, the red dotted lines represent the thin-bar sample in 
Fig.~\ref{fig:Xmanual}. The black dashed line in (c) shows $\langle K_r \rangle$
for a thin bar with perfect FCC order. The solid vertical lines 
in panels (a)-(c) give approximate boundaries between the four regimes 
of vibrational response as a function of cooling rate $R$.}
\label{fig:Xdata}
\end{center}
\end{figure}

Snapshots of the zero-temperature thin-bar resonators generated using
six different cooling rates are shown in Fig.~\ref{fig:Xdefects} (with 
periodic boundary conditions and prior to adding excitations).  We
use the Common Neighbor Analysis (CNA)~\cite{honeycutt1987molecular} to
identify atoms that occur in crystalline (either face-centered cubic
(FCC) or hexagonal close packed (HCP)) and amorphous environments in
the thin bars.  In Fig.~\ref{fig:Xdata} (a), we show that the
ensemble-averaged fraction of crystalline atoms $\langle f_X \rangle$
decreases with increasing $R$. $\langle f_X \rangle$ is nearly $90\%$
when $R=1.2 \times 10^{-4}$ and $\langle f_X \rangle = 0$ for $R=1.2
\times 10^{-1}$. Near the critical cooling rate $R_c \approx
10^{-2.5}$, the system contains a roughly equal mixture of atoms in
crystalline and amorphous environments.

To quantify disorder in the thin-bar samples, we used the Dislocation
Extraction Analysis (DXA) tool within the OVITO software
library~\cite{stukowski2009visualization}. DXA allows us to measure
the total dislocation length $L_D$, which gives the sum of the
magnitudes of the Burgers vectors for each dislocation in the sample.
For $R \ll 1$, we expect few detects, and thus $\langle L_D \rangle
\rightarrow 0$.  In Fig.~\ref{fig:Xdata} (b), for small $R$, we show
that $L_D$ increases with cooling rate $R$~\cite{blanter2007internal,nowick2012anelastic}.  When $R> 3 \times
10^{-3}$, $\langle L_D \rangle$ drops sharply since the thin-bar
samples include mixtures of atoms with crystalline and amorphous
environments. $L_D \rightarrow 0$ when the sample becomes completely
amorphous.
 
To determine the vibrational response, we excite the fundamental mode
$\omega_1$ for each sample, and measure $\langle K_r \rangle$ as a
function of $R$. The behavior for $\langle K_r \rangle$ can be divided
into four regimes.  (See Fig.~\ref{fig:Xdata} (c).) First, at low
cooling rates $R \lesssim 3 \times 10^{-3}$ (regime I), the systems
are mostly crystalline with sparse dislocations.  In this regime, as
$R$ increases, more dislocations are formed and $\langle L_D\rangle$
increases, which causes $\langle K_r \rangle$ to decrease.  In regime
II, at intermediate cooling rates $3 \times 10^{-3} \lesssim R
\lesssim 1.2 \times 10^{-2}$, $\langle f_X \rangle$ drops sharply and
the thin bars contain mixtures of crystalline and amorphous atoms. The
additional boundaries between amorphous and crystalline regions of the
system causes a larger decrease in $\langle K_r \rangle$ than at
smaller $R$.  In regime III, $1.2 \times 10^{-2} \lesssim R \lesssim
3 \times 10^{-1}$, the thin-bar resonators become homogeneously amorphous and
metastable, causing $\langle K_r \rangle$ to increase by a factor of
$\approx 4$. At the high cooling rates $R \gtrsim 3 \times 10^{-1}$ in regime IV, $K_r$
will decrease modestly with increasing $R$. For the Kob-Andersen
bidisperse mixture, we already showed in Fig.~\ref{fig:scaling} (a)
that $\langle K_r \rangle$ decreases by a factor of $\approx 3$ as $R$
is increased over three orders of magnitude. This local maximum in
$\langle K_r \rangle(R)$ is interesting because it shows that there is a
regime where amorphous resonators can have larger $Q$-values than
partially crystalline resonators.

\begin{figure}
\begin{center}
\includegraphics[width=0.32\columnwidth]{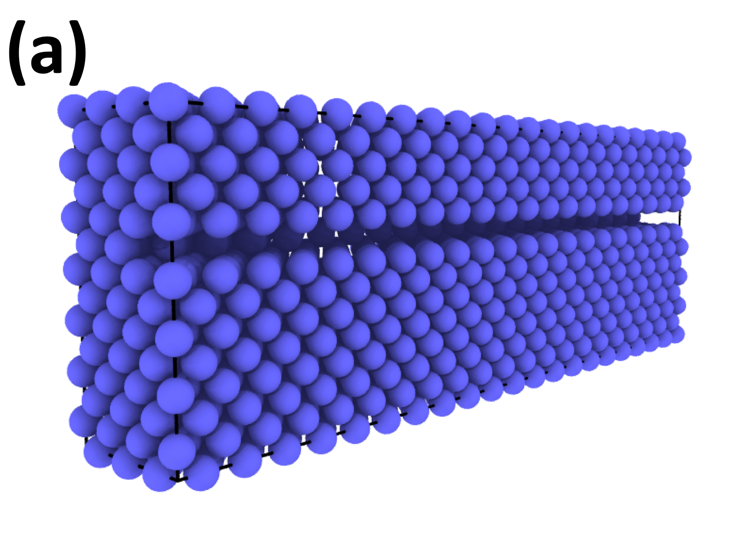}
\includegraphics[width=0.32\columnwidth]{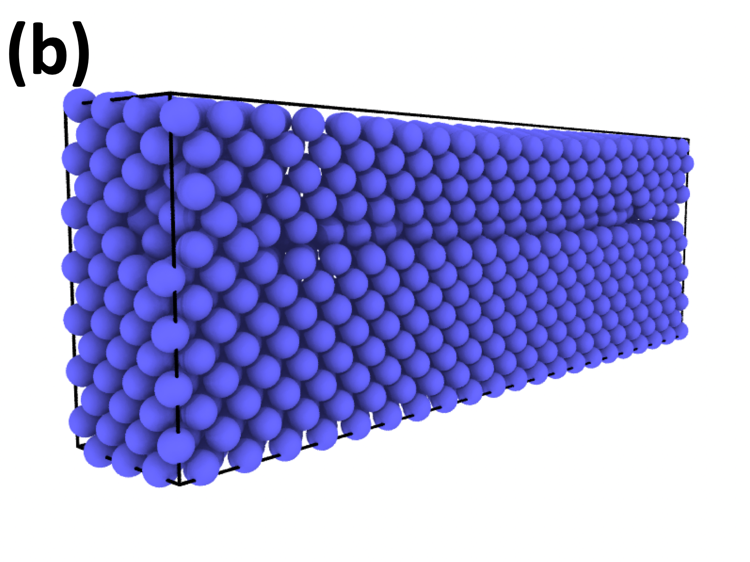}
\includegraphics[width=0.32\columnwidth]{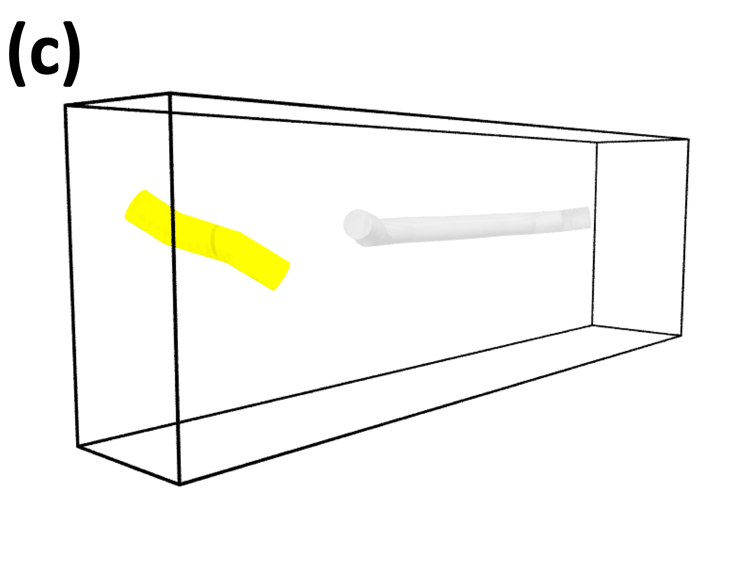}
\caption{Snapshots of thin-bar crystalline resonators with a specifically 
placed defect. In (a), 
we delete a row of atoms from a resonator with perfect FCC order and perform 
energy minimization, which yields the thin-bar resonator in (b). (c) We 
identify two dislocations in the resonator in (b) colored yellow and gray.}
\label{fig:Xmanual}
\end{center}
\end{figure}

In addition to studying the vibrational response of thin-bar
resonators generated by cooling a high-temperature liquid into a
solid, we also investigated the vibrational response of systems for
which we started with perfect FCC crystalline thin bars and generated
specifically placed defects.  In particular, we generated four
thin-bar samples that were initialized with perfect FCC order, and
then we removed a slot with a width of one atom, depth of two atoms,
and varying lengths along different directions in the sample. An
example is shown in Fig.~\ref{fig:Xmanual}. We display $\langle L_D
\rangle$ and $\langle K_r \rangle$ for these four systems in
Fig.~\ref{fig:Xdata} (b) and (c). These samples possess a range of
$\langle K_r \rangle$: some values are larger than that of the rapidly
cooled glass ($R=1.2 \times 10^{-1}$), while others are not.  These
results show that amorphous resonators can possess values of $\langle
K_r \rangle$ (and thus $Q$) that are comparable to those for
crystalline samples.  For example, the thin-bar resonator
corresponding to the green dotted horizontal line in
Fig.~\ref{fig:Xdata} (c) possesses a smaller $\langle K_r \rangle$ than that of the
rapidly cooled glass, with a dislocation density $\langle L_D\rangle
/V \approx 2\times 10^{16}$~m$^{-2}$, which is similar to the value
for crystalline metals with strong
dislocations~\cite{schafler2001measurement}.  Since metallic glasses
do not need to be annealed, can be molded into complex shapes, possess
unique magnetic and biocompatibility properties~\cite{li2012amorphous,schroers2009bulk,schroers2010processing,kumar2009nanomoulding,li2018atomic}, and can possess comparable
quality factors to crystalline metals~\cite{kanik2014high}, metallic
glasses are promising materials for high-$Q$ applications.

\section{conclusion} 
\label{sec:conclusion}

In this article, we employ molecular dynamics simulations of model
metallic glass resonators undergoing vibrations to quantify the
intrinsic dissipation and loss mechanisms caused by thermal
fluctuations and atomic rearrangements. Using thin-bar resonators
generated over a wide range of cooling rates, we excite the
fundamental mode corresponding to the lowest eigenfrequency $\omega_1$
of the dynamical matrix as a function of the kinetic
energy per atom $K$.  We find three regimes of vibration. In the
linear response regime, $K<K_{nl}$, most of the energy of the
vibrations remains in the fundamental mode, the loss is small, and $Q
\rightarrow \infty$ (since we do not consider coupling of the 
resonator to the
environment).  For $K_{nl}<K<K_r$, energy can leak from the
fundamental mode to others at short times, but at sufficiently long
times the leakage of energy to other frequencies stops, and thus $Q \rightarrow
\infty$ at long times.  For $K > K_r$, one or more atomic
rearrangements occur.  In this regime, energy in the fundamental
mode is completely redistributed to a large set of other frequencies,
the loss is large, and $Q$ is finite. Thus, we show that $K_r$ determines
the quality factor. 

We find that $\langle K_r \rangle$ decreases as a power-law
$N^{-k}$ with increasing system size $N$, where $k \approx 1.3$ decreases
only modestly with decreasing $R$. We find similar results for the critical
shear strain $\langle K_r \rangle \sim \langle \gamma_r \rangle ^2$
using athermal quasistatic shear deformation, where $\langle \gamma_r
\rangle$ is the characteristic strain above which atomic
rearrangements begin to occur. Using these results, we estimate that
$\langle \gamma_r \rangle \sim 10^{-8}$ for micron-sized resonators,
and thus large $Q$-values can be obtained when these resonators are
operated at $\gamma < \gamma_r$. We also measured $\langle K_r \rangle$
in thin-bar resonators with crystalline order and compared the
vibrational response to that in amorphous resonators. We find that
$\langle K_r \rangle$ is similar for amorphous resonators and those
with significant crystalline order. In light of the fact that metallic
glasses can be thermoplastically formed into complex shapes, possess
unique magnetic and biocompatibility properties, and can achieve
$Q$-values that are comparable to those for crystalline structures,
metallic glasses are promising materials for micro- and
nano-resonators.

Our results raise a number of interesting future directions.  For
example, we can investigate methods that involve mechanical
deformation, not slower cooling rates or annealing methods, to
increase $\langle K_r \rangle$ and move the sample to regions of
configuration space with higher energy barriers between inherent
structures. One possible approach is to apply athermal cyclic simple
or pure shear deformation to samples that have been prepared using
fast cooling rates. Recent studies have found that there is a finite
critical strain amplitude for cyclic shear that marks the limit
between reversible and irreversible atomic rearrangements in the
large-system limit~\cite{regev2015reversibility,fiocco2014encoding,leishangthem2017yielding}.  Does this imply
that cyclic shear training can find zero-temperature configurations
for which $\langle K_r \rangle$ remains finite in the large-system
limit?  In addition, we can explore how the type of cyclic driving
affects $\langle K_r \rangle$ and whether configurations can be
trained in multiple directions simultaneously to increase $\langle K_r
\rangle$. Another future direction involves studies of the loss and quality 
factor when the resonator has clamped instead of free boundary conditions 
and when it is driven over a range of frequencies, not only the 
fundamental mode. 

% If in two-column mode, this environment will change to single-column format so that long equations can be displayed. 
%\begin{widetext}
%$$\mbox{put long equation here}$$
%\end{widetext}

% Figures should be put into the text as floats. 
% Use the graphics or graphicx packages (distributed with LaTeX2e).
% Put the label that you will use with \ref{} command in the braces of the \label{} command.
%
% \begin{figure}
% \end{figure}

% \begin{table}
% \end{tabular}
% \end{table}

\begin{acknowledgments}
% Put your acknowledgments here.
The authors acknowledge support from NSF MRSEC Grant No. DMR-1119826 (M. F. and C.O.) and NSF Grant Nos. CMMI-1462439 (M. F. and C.O.) and CMMI-1463455 (M.S.). This work was supported by the High Performance Computing facilities operated by, and the staff of, the Yale Center for Research Computing.
\end{acknowledgments}

\appendix

\section{Length of time series}
\label{sec:appendix_tapelength}

\begin{figure}
\begin{center}
\includegraphics[width=0.96\columnwidth]{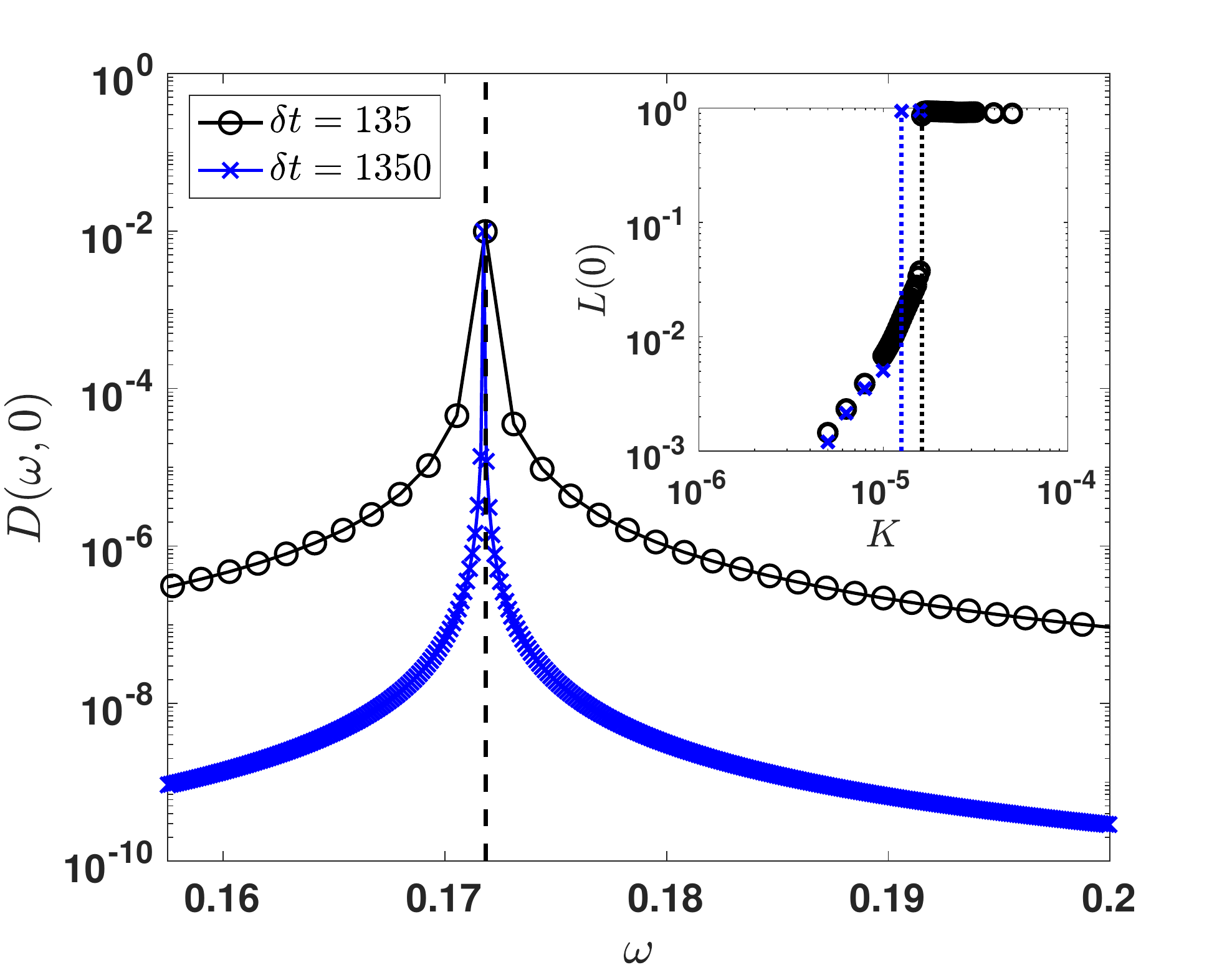}
\caption{The low-frequency regime for the density of 
vibrational modes $D(\omega,0)$ near the fundamental frequency $\omega_1 
\approx 0.172$ (vertical dashed line) calculated using time series with two 
different total lengths, $\delta t=135$ (black circles) and $1350$ (blue exes),
at $K=5\times10^{-6}$ in the linear response regime. In the inset, we 
show the loss $L(0)$ versus $K$ for the same time series in the main panel.
The characteristic kinetic energy per atom above which an atomic 
rearrangement occurs are indicated: $K_r=1.62\times10^{-6}$
(black dotted line) for $\delta t=135$
and $K_r=1.26\times10^{-6}$ (blue dotted line) for $\delta t=1350$.
}
\label{fig:tapelength}
\end{center}
\end{figure}

In the main text, we used a total run length of $\delta t=135$ to
calculate the density of vibrational modes and loss for the first time
interval $t=0$ in Figs.~\ref{fig:spec} and~\ref{fig:loss} (as well as 
all other time intervals). In this
Appendix, we show results for $D(\omega,0)$ when $\delta t$ is
increased by a factor of $10$ (keeping the sampling rate fixed).  In
Fig.~\ref{fig:tapelength}, for $K<K_r$ in the linear response regime,
we show that the peak value of $D(\omega_1,0)$ is unchanged for
$\delta t = 135$ and $1350$, and thus $L(0)$ is nearly the same
for the two values of $\delta t$.  We know that the probability for an atomic
rearrangement increases with time $\delta t$ at fixed $K$. Thus, in the
inset to Fig.~\ref{fig:tapelength}, we show that the loss $L(0)$
undergoes a discontinuous jump for $\delta t = 1350$ at a smaller $K$
than that for $\delta t = 135$. We find that $K_r \approx
1.62\times10^{-5}$ for $\delta t=135$ and $\approx
1.26\times10^{-5}$ for $\delta t=1350$.  Thus, the precise value of 
$K_r$ depends on $\delta t$, but all of the results are qualitatively the 
same for different choices of $\delta t$. 

\section{Time-dependent density of vibrational modes}
\label{sec:appendix_dissipating}

\begin{figure*}
\begin{center}
\includegraphics[width=0.68\columnwidth]{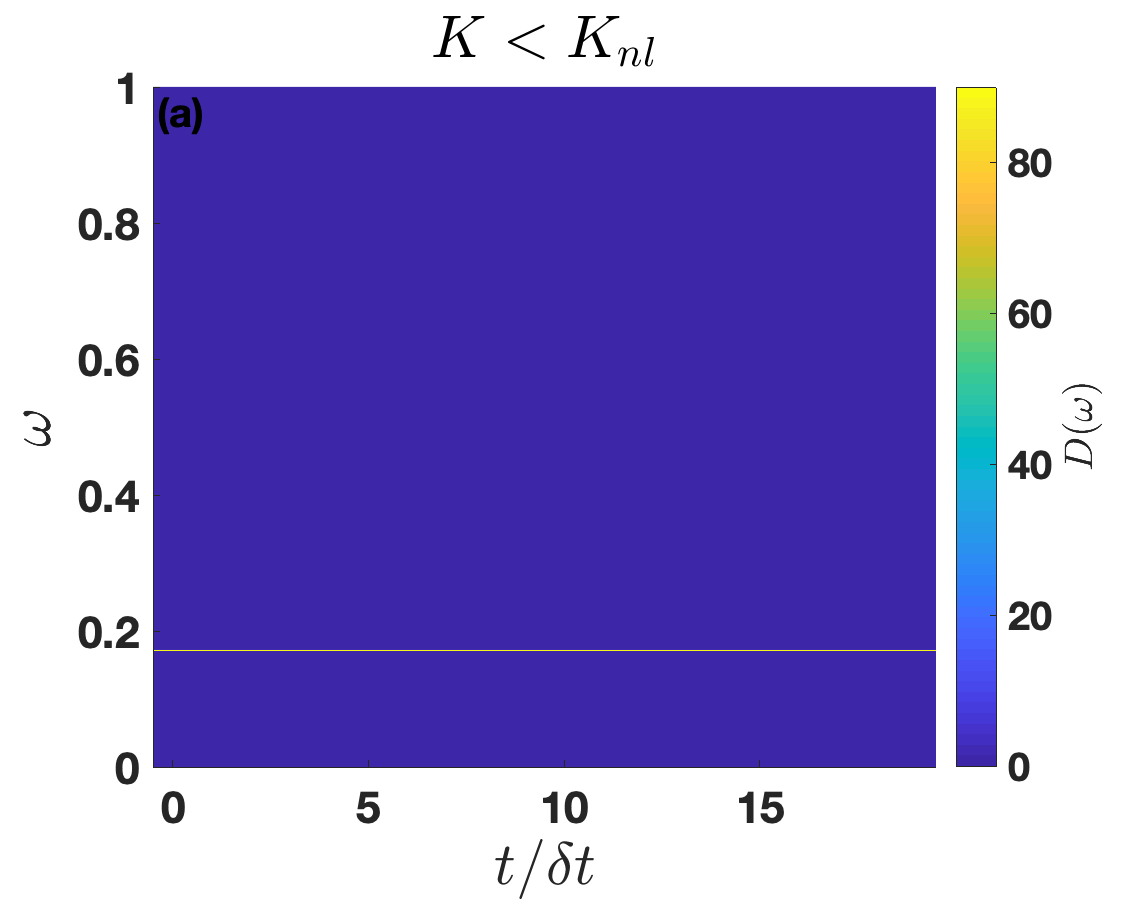}
\includegraphics[width=0.68\columnwidth]{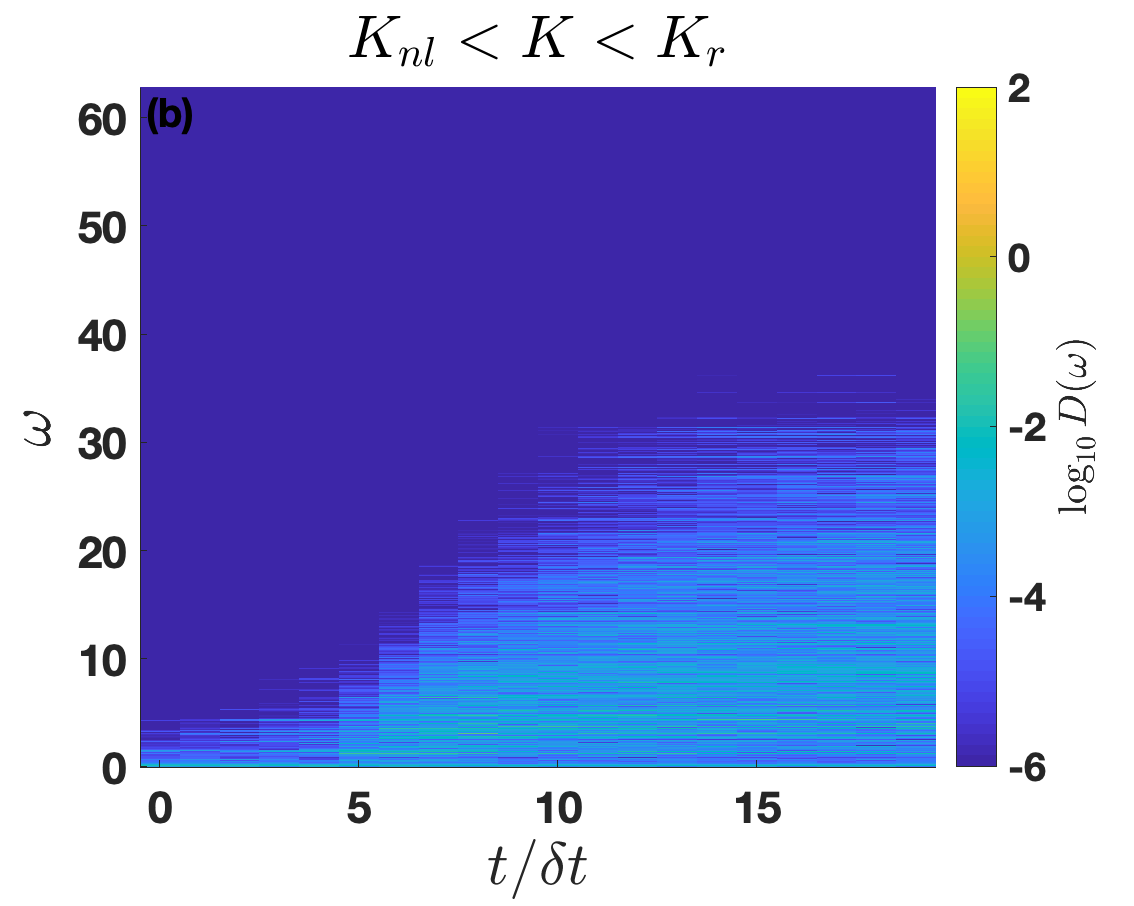}
\includegraphics[width=0.68\columnwidth]{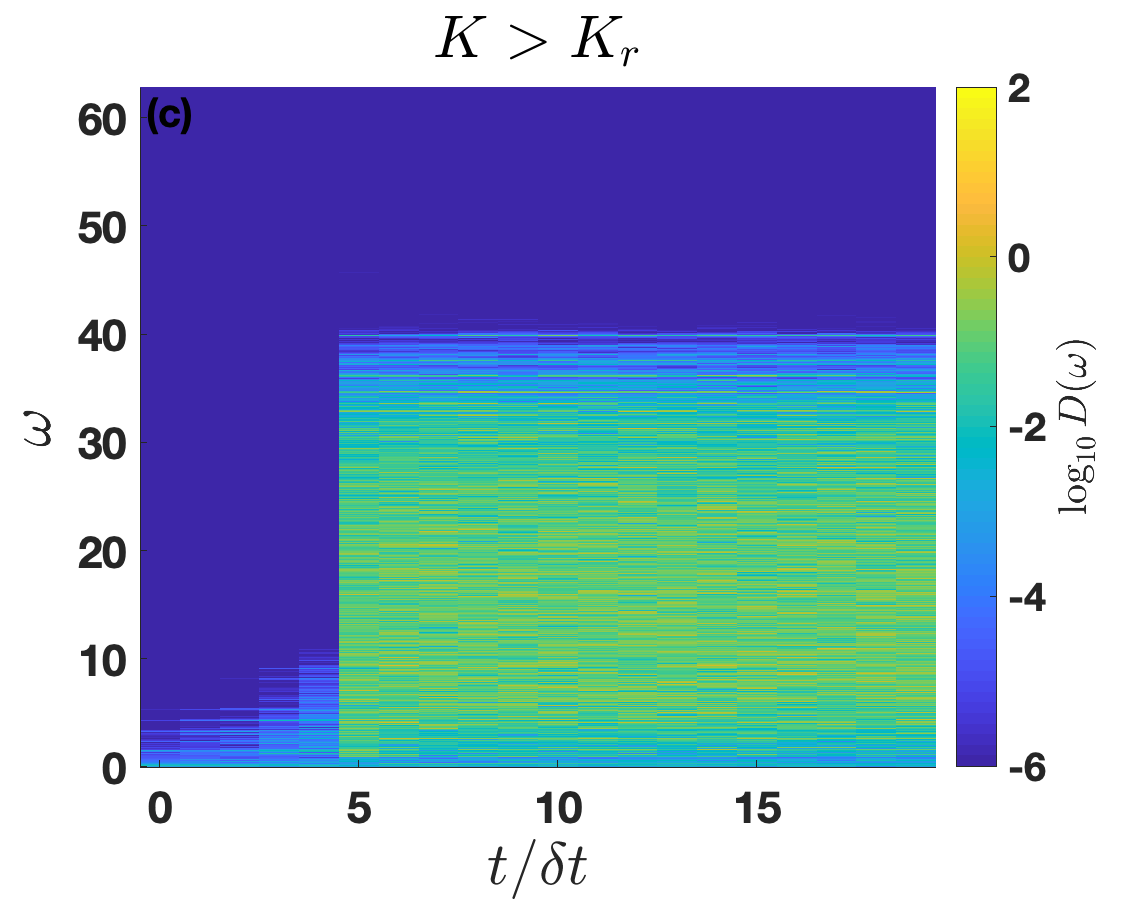}
\caption{The density of vibrational modes $D(\omega,t)$ versus 
time $t/\delta t$ in the three regimes for the kinetic energy per atom,  
(a) $K < K_{nl}$, (b) $K_{nl} < K < K_r$, and (c) $K > K_r$, for a $N=2000$ thin-bar resonator with amorphous structure. The color scale from yellow to 
dark blue give decreasing values of $D(\omega,t)$ in panel (a), and $\log_{10}{D(\omega,t)}$
in panels (b) and (c).}
\label{fig:dissipating}
\end{center}
\end{figure*}

For the calculations of the density of vibrational modes and loss in
the main text, we divided a long time series following the excitation
of the resonator along the fundamental mode into $20$ time intervals
of equal length $\delta t$. We showed the density of vibrational modes
$D(\omega,0)$ for the first time interval (i.e. considering times from
$0$ to $\delta t$) in Fig.~\ref{fig:spec}. In this Appendix, we
calculate $D(\omega,t)$ for all $20$ time intervals. We show
$D(\omega,t)$ for $K$ values in the three regimes, $K < K_{nl}$,
$K_{nl} < K < K_r$, and $K > K_r$, which match those used in
Fig.~\ref{fig:q} (a).  For $K<K_{nl}$, there is minimal leakage of
energy from the fundamantal mode $\omega_1 = 0.172$ and $Q \rightarrow
\infty$.  In the regime $K_{nl}<K<K_r$, energy leaks from the
fundamental mode at short times, but it stops for $t/\delta t \gtrsim
12$, and the system vibrates nonlinearly with finite loss, finite $Q$
for $t/\delta t \lesssim 12$ and $Q\rightarrow \infty$ for $t/\delta t
\gtrsim 12$. For $K>K_r$, strong energy leakage occurs due to an
atomic rearrangement at $t/\delta t \approx 5$.

%\section*{references}
\bibliography{resonator}

\end{document}